%% file: sample-acmsmall-conf.tex
\documentclass[acmsmall]{acmart}

\input{meta/lstcustom}
\input{meta/algorithmstyle}
\input{meta/package}

\newcommand{\blackding}[1]{\ding{\numexpr181+#1\relax}}
\newcommand{\whiteding}[1]{\ding{\numexpr171+#1\relax}}

\newcommand{\mybox}[1]{\begin{tcolorbox}[enhanced, frame hidden, boxsep=0pt]\textnormal{#1}\end{tcolorbox}}

\newcommand{\keenhash}{\text{KEENHash}}  
\newcommand{\keenhashsemantic}{\text{KEENHash$_\text{sem}^f$}}
\newcommand{\keenhashsemanticbinary}{\text{KEENHash$_\text{sem}^b$}}
\newcommand{\keenhashstructural}[1]{\text{KEENHash$_\text{stru}^{#1}$}}
\newcommand{\psso}{$\text{PSS}_\text{O}$}

\newcommand{\minhashstring}{Minhash$_\text{s}$}
\newcommand{\meanpooling}{\text{$\text{Mean Pooling}$}}

\newcommand{\pythia}{\texttt{Pythia-410M}}

\newcommand{\revision}[1]{\textcolor{black}{#1}}

\AtBeginDocument{%
  }

\setcopyright{cc}
\setcctype{by}
\acmDOI{10.1145/3728911}
\acmYear{2025}
\acmJournal{PACMSE}
\acmVolume{2}
\acmNumber{ISSTA}
\acmArticle{ISSTA036}
\acmMonth{7}
\received{2024-10-30}
\received[accepted]{2025-03-31}

\begin{document}


\title[KEENHash: Hashing Programs into Function-Aware Embeddings...]{KEENHash: Hashing Programs into Function-Aware Embeddings for Large-Scale Binary Code Similarity Analysis}


\author{Zhijie Liu}
\orcid{0000-0003-0972-8631}
\affiliation{%
  \institution{ShanghaiTech University}
  \city{Shanghai}
  \country{China}
}
\email{liuzhj2022@shanghaitech.edu.cn}

\author{Qiyi Tang}
\orcid{0000-0002-8200-7518}
\affiliation{%
  \institution{Tencent Security Keen Lab}
  \city{Shanghai}
  \country{China}
}
\email{work_t71@163.com}

\author{Sen Nie}
\orcid{0000-0003-4154-2941}
\affiliation{%
  \institution{Tencent Security Keen Lab}
  \city{Shanghai}
  \country{China}
}
\email{snie@tencent.com}

\author{Shi Wu}
\orcid{0000-0002-6842-7487}
\affiliation{%
  \institution{Tencent Security Keen Lab}
  \city{Shanghai}
  \country{China}
}
\email{shiwu@tencent.com}

\author{Liang Feng Zhang}
\orcid{0000-0003-3543-1524}
\affiliation{%
  \institution{ShanghaiTech University}
  \city{Shanghai}
  \country{China}
}
\email{zhanglf@shanghaitech.edu.cn}

\author{Yutian Tang}
\authornote{Yutian Tang (yutian.tang@glasgow.ac.uk) is the corresponding author.}
\orcid{0000-0001-5677-4564}
\affiliation{%
  \institution{University of Glasgow}
  \city{Glasgow}
  \country{United Kingdom}
}
\email{yutian.tang@glasgow.ac.uk}


\input{secs/abstract}



\maketitle

\input{secs/introduction}
\input{secs/preliminary}

\input{secs/methodology}
\input{secs/evaluation}

\input{secs/discussion}

\input{secs/relatedwork}

\input{secs/conclusion}



\begin{acks}
We thank Dr. Pengfei Jing for the helpful discussion when preparing the manuscript. This work is partially supported by the National Natural Science Foundation of China (Grant No. 62202306 and Grant No. 62372299). Zhijie Liu would like to dedicate this paper to the love of his fiancée.
\end{acks}

\bibliographystyle{ACM-Reference-Format}
\bibliography{refs/refs}

\newpage
\appendix
\input{secs/appendix}

\end{document}

%% file: meta/lstcustom.tex
\usepackage{color}
\usepackage{etoolbox}
\usepackage{listings}
\usepackage[T1]{fontenc}

\newcommand{\lstfontfamily}{\ttfamily}

\definecolor{darkviolet}{rgb}{0.5,0,0.4}
\definecolor{darkgreen}{rgb}{0,0.4,0.2} 
\definecolor{darkblue}{rgb}{0.1,0.1,0.9}
\definecolor{darkgrey}{rgb}{0.5,0.5,0.5}
\definecolor{lightblue}{rgb}{0.4,0.4,1}

\definecolor{stringColor}{rgb}{0.16,0.00,1.00}
\definecolor{annotationColor}{rgb}{0.39,0.39,0.39}
\definecolor{keywordColor}{rgb}{0.50,0.00,0.33}
\definecolor{commentColor}{rgb}{0.25,0.50,0.37}
\definecolor{javadocColor}{rgb}{0.25,0.37,0.75}
\definecolor{jTagColor}{rgb}{0.50,0.62,0.75}
\definecolor{eTagColor}{rgb}{0.50,0.62,0.75}
\definecolor{lineNumberColor}{rgb}{0.47,0.47,0.47}

\def\jTags{@author, @deprecated, @exception, @param, @return, @see, @serial, @serialData, @serialField, @since, @throws, @version}

\def\jAnnotations{
    classoffset=1,
    morekeywords={@Override, @Deperecated, @SuppressWarnings, @Retention, @Documented, @Target, @Inherited},
    keywordstyle=\color{annotationColor},
    classoffset=0
}

\def\eTags{FIXME, TODO, XXX}

\newrobustcmd{\markupJavadocs}[1]{%
\edef\mytok{\the\lst@token}%
\renewcommand*{\do}[1]{%
\ifdefstring{\mytok}{##1}%
{\color{jTagColor}\bfseries\listbreak}%
{}%
}%
{\color{javadocColor}%
\expandafter\docsvlist\expandafter{\jTags}%
\renewcommand*{\do}[1]{%
\ifdefstring{\mytok}{##1}%
{\color{eTagColor}\bfseries\listbreak}%
{}%
}%
\expandafter\docsvlist\expandafter{\eTags}%
#1}%
}%

\newrobustcmd{\markupComments}[1]{%
\edef\mytok{\the\lst@token}%
\renewcommand*{\do}[1]{%
\ifdefstring{\mytok}{##1}%
{\color{eTagColor}\bfseries\listbreak}%
{}%
}%
{\color{commentColor}%
\expandafter\docsvlist\expandafter{\eTags}#1}%
}%

\lstdefinestyle{eclipse}{
  basicstyle={\lstfontfamily},
  emphstyle=\bfseries,
  keywordstyle=\color{keywordColor}\bfseries,
  commentstyle=\markupComments,
  stringstyle=\color{stringColor},
  numberstyle=\color{lineNumberColor}\lstfontfamily,
  morecomment=[s][\markupJavadocs]{/**}{*/}, 
  showstringspaces=false,
  numbers=left,
}

\lstdefinestyle{black}{
  basicstyle=\small\lstfontfamily,
  numbers=left,
  columns=fullflexible,
  breaklines=true,
  mathescape=true,
  escapechar=\#,
  tabsize=4,
  frame=lines,
  showstringspaces=false
}

\lstdefinestyle{seminar}{
  basicstyle=\small\ttfamily,
  numbers=left,
  breaklines=true,
  mathescape=true,
  escapechar=\#,
  tabsize=4,
  showstringspaces=false
}

\expandafter\lstset\expandafter{\jAnnotations}

%% file: meta/algorithmstyle.tex
\usepackage{algorithm2e}
\usepackage{xcolor}

\SetAlFnt{\footnotesize}

\SetCommentSty{xCommentSty}

\LinesNumbered
\SetSideCommentRight
\DontPrintSemicolon
\RestyleAlgo{algoruled}

\usepackage{siunitx}
\usepackage[autolanguage]{numprint}

\newcommand*\np[2][z]{
\ifx z#1%
$\numprint{#2}$%
\else%
$\numprint[#1]{#2}$%
\fi\xspace%
}

\usepackage{fp}
\newcommand{\ShowAbsoluteNumber}[1]{%
\ifnum #1<10%
{\hspace*{0pt}#1}%
\else%
\ifnum #1<100%
{\hspace*{0pt}#1}%
\else%
\ifnum #1<1000%
{\hspace*{0pt}#1}%
\else%
{\numprint{#1}}%
\fi%
\fi%
\fi%
}

\newcommand{\ShowPercentage}[2]{%
\FPeval\percentage{round(#1/#2*100,0)}%
\FPeval\percentageOneDecimal{round(#1/#2*100,1)}%
\ifnum \percentage=0%
{\np[\%]{\FPprint{percentageOneDecimal}}}%
\else%
\ifnum \percentage<10%
{\np[\%]{\FPprint{percentageOneDecimal}}}%
\else%
{\np[\%]{\FPprint{percentageOneDecimal}}}%
\fi%
\fi%
\xspace
}
\newlength\BARSIZE  
\newcommand{\inlinechart}[2]{%
\FPeval{\BLACKBARSIZE}{#1/#2}\textcolor{black!80}{\rule{\BLACKBARSIZE\BARSIZE}{1.6ex}}%
\FPeval{\BLACKBARSIZE}{1 - (#1/#2)}\textcolor{black!10}{\rule{\BLACKBARSIZE\BARSIZE}{1.6ex}}%
}

\newcommand*\percent[3][v]{%
\ifx q#1%
    \np{#2}/\np{#3}(\ShowPercentage{#2}{#3})\else%
\ifx p#1%
    \np{#2}(\ShowPercentage{#2}{#3})\else%
\ifx c#1%
    \inlinechart{#2}{#3}%
\else%
    \np{#2}%
    \ifx r#1%
        /\np{#3}%
    \fi%
    \hspace*{0.5ex}(\ShowPercentage{#2}{#3}) %
    \inlinechart{#2}{#3}%
    \xspace
\fi\fi\fi%
}

%% file: meta/package.tex
\usepackage{hyperref}
\usepackage{color}
\usepackage[most]{tcolorbox}
\usepackage{balance}
\usepackage{makecell}
\usepackage{verbatim}
\usepackage{mdframed}
\usepackage{lipsum}
\usepackage{amsmath}
\usepackage{amssymb}
\usepackage{amsfonts}
\usepackage[font=small]{caption}
\usepackage{ifpdf}
\usepackage{graphicx}
\usepackage{wrapfig}

\usepackage{pifont}

\usepackage[inkscapelatex=false]{svg}

\usepackage{algorithmic}
\usepackage{array}

\usepackage{caption}
\usepackage{multirow}
\usepackage{float}
\usepackage{mwe}
\newcommand{\code}[1]{\texttt{#1}}

\usepackage{url}
\newcommand{\TBD}[2]{{\color{red}(TBD-#1: #2)}}

\usepackage{subfigure}
\usepackage{tikz}
\usepackage{xcolor}

\usepackage{booktabs,caption}  
\usepackage[flushleft]{threeparttable}

 \usepackage{enumitem}

\hypersetup{
 colorlinks=true,
 linkcolor=blue,
 citecolor=blue,
 filecolor=black,
 urlcolor=black,
 pdfauthor = {},
 pdftitle = {},
 pdfkeywords = {}
}

%% file: secs/abstract.tex
\begin{abstract}
Binary code similarity analysis (BCSA) is a crucial research area in many fields such as cybersecurity.
%
%
Specifically, function-level diffing tools are the most widely used in BCSA: they perform function matching one by one for evaluating the similarity between binary programs. However, such methods need a high time complexity, making them unscalable in large-scale scenarios (e.g., $1$/$n$-to-$n$ search).
%
%
Towards effective and efficient program-level BCSA, we propose \keenhash{}, a novel hashing approach that hashes binaries into program-level representations through large language model (LLM)-generated function embeddings. \keenhash{} condenses a binary into one compact and fixed-length program embedding using K-Means and Feature Hashing, allowing us to do effective and efficient large-scale program-level BCSA, surpassing the previous state-of-the-art methods.
%
%
%
The experimental results show that \keenhash{} is \revision{at least} 215 times faster than the state-of-the-art function matching tools while maintaining effectiveness. Furthermore, in a large-scale scenario with 5.3 billion similarity evaluations, \keenhash{} takes only 395.83 seconds while these tools will cost \revision{at least} 56 days.
We also evaluate \keenhash{} on the program clone search of large-scale BCSA across extensive datasets in 202,305 binaries. 
%
Compared with 4 state-of-the-art methods, \keenhash{} outperforms all of them by at least 23.16\%, and displays remarkable superiority over them in the large-scale BCSA security scenario of malware detection.
\end{abstract}

%% file: secs/introduction.tex
\section{Introduction}\label{sec:introduction}

Binary code similarity analysis (BCSA) is a crucial research area in the fields of cybersecurity, software engineering, and reverse engineering~\cite{benoit2023scalable, ding2019asm2vec, jiang2024binaryai, hu2013mutantx, duan2020deepbindiff, kim2022revisiting, yu2020order, pal2024len}. It involves the comparison of binary code (e.g., program and function) to identify similarities and differences among them, which is applied to a wide range of applications, including code clone search~\cite{ding2019asm2vec, benoit2023scalable, wang2022jtrans}, malware analysis~\cite{hu2013mutantx, jang2011bitshred, duan2020deepbindiff, dambra2023decoding, benoit2023scalable, virustotal}, vulnerability detection~\cite{wang2023sem2vec, feng2016scalable, xu2017neural}, software composition analysis~\cite{jiang2023third, jiang2024binaryai, yuan2019b2sfinder}, and so forth. 
Among these analyses, program-level BCSA~\cite{haq2021survey} stands out as a powerful technique that can analyze and compare similarities between binaries (i.e., binary programs)~\cite{benoit2023scalable, virustotal, hu2013mutantx} which are larger and more complex objects than functions (i.e., function-level BCSA, evaluates the similarity between binary functions). This type of analysis includes $1$-to-$1$ and $1$/$n$-to-$n$ similarity comparisons of binaries. Particularly, in the \textbf{vital large-scale scenarios} (i.e., $1\text{/}n$-to-$n$), such as program clone search~\cite{benoit2023scalable, hu2013mutantx}, malware detection~\cite{dambra2023decoding, virustotal}, and threat intelligence~\cite{virustotal, li2019reading, binaryai, bouwman2020different, mink2023everybody}, it must be both accurate and efficient for evaluating huge amounts of similarities among binaries. 

Unfortunately, \revision{in spite} of the significance on program-level BCSA, none of the previous works could achieve satisfying results on the large-scale $1$/$n$-to-$n$ comparisons.
Specifically, most previous works on BCSA ~\cite{benoit2023scalable, ding2019asm2vec, wang2022jtrans, wang2024clap, duan2020deepbindiff} focus on \emph{function-level} similarity: they generate one embedding for each function in a binary, and iteratively compare these embeddings with embeddings from another binary, to determine the (similar) function-matched proportion (i.e., similarity) between two binaries. \textbf{\textit{Limitation.1:}} Such a matching approach has an unscalable $O(nm^3)$ time complexity ($n$ is the number of comparisons among binaries and $m$ is the number of functions to one binary)~\cite{kuhn1955hungarian}, making it impossible to be applied to large-scale BCSA, even with faster heuristic strategy~\cite{binaryai, bindiffmatchgithub, duan2020deepbindiff} (Sec. \ref{sec:eval:functionmatching}). \textbf{\textit{Limitation.2:}} \psso{}~\cite{benoit2023scalable} is the only recent work focusing on large-scale program-level BCSA. However, it only uses simple features (e.g., call graph and edge counts of control flow graphs) for generating program-level embedding without considering the rich semantics in binary functions, leading to poor performance in large-scale experiments (Sec. \ref{sec:eval:programclonesearch} to \ref{sec:eval:malware}). 

To fill the \revision{aforementioned} gaps, we propose \keenhash{}, a novel hashing approach to hash binaries into fixed-length representations for \emph{effective and efficient} BCSA.
To achieve this goal, \keenhash{} condenses a binary into one compact and fixed-length program embedding (e.g., 8KB).
\revision{It leverages} K-Means and Feature Hashing technique~\cite{weinberger2009feature} \revision{to classify and represent decompiled pseudo functions (with function embeddings) into a bit-vector for approximating function matching}, \revision{thereby} allowing us to do large-scale ($1$-to-$n$ and $n$-to-$n$) program-level BCSA that previous function-level methods fail to do.
Specifically, \keenhash{} involves three stages to hash a binary: Function Embedding Generation (Sec. \ref{sec:functionembeddinggeneration}), Program Embedding Generation (Sec. \ref{sec:programembeddinggeneration}), and Similarity Evaluation (Sec. \ref{sec:similarityevaluation}).
In the first stage, we train a large language model (LLM) to encode the decompiled pseudocode of each function (i.e., pseudo function) into corresponding embedding, in which the rich semantics of pseudo functions are extracted and maintained.
Taking the idea of classifying similar functions into the same class and thus representing function matching, in the second stage, we cluster extensive source functions using K-Means (an effective clustering algorithm to generate labels), and then feed the label information (derived from K-Means through classification) of the pseudo functions into the Feature Hashing module to produce the compact vector (\keenhash{}-stru, Sec. \ref{sec:structurebasedembeddinggeneration}) that can represent the whole binary.
By applying such K-Means and Feature Hashing techniques, an original binary with the size of MBs of pseudo function embeddings can be condensed into only 8KB, significantly accelerating the (function matching) process of large-scale program-level BCSA.
Furthermore, massive code reuse (e.g., > 70\%), a widespread practice in software development~\cite{jiang2023third}, can misleadingly increase the similarity between non-same-class binaries (Sec. \ref{sec:eval:programclonesearch}) and weaken indirectly the significance of binaries' unique feature parts. We introduce \keenhash{}-sem (Sec. \ref{sec:semanticsbasedembeddinggeneration}) from the perspective of program semantics with the weighted average of function embeddings.
Comparing these two approaches, \keenhash{}-stru is better in code obfuscation scenarios (Sec. \ref{sec:eval:againstobfuscation}), while \keenhash{}-sem is more effective in massive code reuse ones (Sec. \ref{sec:eval:programclonesearch}).
Finally, in the third stage, we use the compact program embedding (\keenhash{}-stru or sem) representing the information of the whole binary to perform the large-scale BCSA.

Our experiments illustrate that \keenhash{} can maintain an effective performance on function matching while faster than \revision{SigmaDiff~\cite{gao2024sigmadiff} and} BinDiffMatch~\cite{bindiffmatchgithub, binaryai}, the two state-of-the-art binary diffing tools for function matching with heuristic strategy, by \revision{1,254 and} 215 times on the average of each matching between two binaries (Sec. \ref{sec:eval:functionmatching}). Furthermore, in a large-scale scenario with 5.3 billion similarity evaluations among binaries (Sec. \ref{sec:eval:all}), \textit{a typical workload within one day}~\cite{virustotal, van2023deep, dambra2023decoding}, \keenhash{} takes at most only 395.83 seconds, while \revision{SigmaDiff} and BinDiffMatch will cost \revision{323 and} 56 days, which is unscalable. Additionally, on program clone search (a representative task for large-scale BCSA), \keenhash{} shows effective performance on a total of 202,305 Linux/Windows binaries with real-world cases across various compile environments including optimization levels, compilers, architectures, and obfuscations, significantly outperforming all other state-of-the-art BCSA methods including \psso{} (Sec. \ref{sec:eval:programclonesearch} to \ref{sec:eval:all}) by at least 23.16\% (Sec. \ref{sec:eval:all}). Moreover, \keenhash{} displays remarkable superiority over other methods in the large-scale BCSA security scenario of malware detection (Sec. \ref{sec:eval:malware}).


To summarize, our paper makes three contributions:

\noindent $\bullet$ We propose \keenhash{}, based on LLM-generated function embeddings, taking two perspectives of function matching and program semantics to hash any binary to a compact and fixed-length representation for effective and efficient large-scale program-level BCSA;


\noindent $\bullet$ We evaluate \keenhash{} on function matching, against the state-of-the-art tools \revision{SigmaDiff and} BinDiffMatch. \keenhash{} is able to maintain an effective performance and demonstrates a significant speed advantage, being \revision{1,254 and} 215 times faster (Sec. \ref{sec:eval:functionmatching}). In a large-scale scenario with 5.3 billion similarity evaluations, it takes at most 395.83 seconds while \revision{SigmaDiff and} BinDiffMatch will cost \revision{323 and} 56 days (Sec. \ref{sec:eval:functionmatching} and \ref{sec:eval:all}), respectively;

\noindent $\bullet$ We evaluate \keenhash{} on program clone search with 5 large-scale datasets in a total of 202,305 Linux/Windows binaries. Experimental results show that \keenhash{} outperforms all other state-of-the-art BCSA methods including \psso{} by at least 23.16\% (Sec. \ref{sec:eval:programclonesearch} - \ref{sec:eval:all}), and displays remarkable superiority over them in the large-scale security scenario of malware detection (Sec. \ref{sec:eval:malware}).

\noindent \textbf{Online SDK Availability for K(EEN)Hash:} \includegraphics[height=0.25cm]{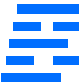} \href{https://www.binaryai.cn}{ \textbf{https://www.binaryai.cn}}.





%% file: secs/preliminary.tex
\section{Preliminary}\label{sec:preliminary}

\subsection{Problem Definition}\label{sec:problemdefinition}

Program-level BCSA is a task to evaluate the similarity between two binary programs. Given two binaries $q$ and $r$, the similarity evaluation process is presented as follows:

\noindent \textbf{Definition 1: (Similarity Evaluation).} For $q$ and $r$, the similarity evaluation process measures the similarity score between them, which is formulated as follows:

\begin{equation}
    \label{equ:similarityevaluation}
\resizebox{0.4\hsize}{!}{$
    Similarity\text{ }Score = F(Enc(q), Enc(r))
$}
\end{equation}

\noindent Where function $Enc$ encodes a program $q$ or $r$ based on their information to a representation; and, function $F$ further measures their similarity score. The larger the similarity score, the more similar the two programs $q$ and $r$ are. This definition also holds between functions.  

In this study, we focus on large-scale program-level BCSA. To evaluate the performance of methods, we utilize the program clone search~\cite{benoit2023scalable} as the evaluation task. \textit{Notably, program clone search is also one of the critical large-scale scenarios in threat intelligence and BCSA (e.g., finding similar binaries to unknown malware for better understanding)}~\cite{virustotal}.

\noindent \textbf{Definition 2: (Same Class Program).}  Programs in the same class are clones of each other. A clone $c$ of a program $p$ is defined as that $c$ is compiled from the same or different code version source code to $p$ with various compilation environments. For example, $c$ compiled from source code $s$ using GCC v13.2 with O0 is a clone of $p$ compiled from $s$ using GCC v10.5 with O3; and, $c$ compiled from $s$ is a clone of $p$ compiled from $s'$ where $s$ is another version to $s'$ (e.g., malware variants).

\noindent \textbf{Definition 3: (Program Clone Search).} Given an unknown query binary $q$ $\in Q$, a query program dataset $Q$, and a program repository dataset $R$ containing a large amount of unknown or known binaries $r$, the task of program clone search is to input $q$ and retrieve the most Top-$k$ similar binaries $\{r_1, r_2, ..., r_k | r_i \in R\}$ from $R$, ranked by their similarity scores. The more binaries of the same class to $q$ are returned and the higher they rank among the Top-$k$ retrieved binaries, the better the performance of program-level BCSA methods.

The clone search procedure is presented as follows:

\noindent \blackding{1} \textbf{Repository Preprocessing.} Before retrieving Top-$k$ similar programs, the repository $R$ needs to be built first. Additionally, for each program $r \in R$, the program-level BCSA method $Enc(r)$ to get its representation, such as embedding, for subsequent similarity comparisons in retrieving;

\noindent \blackding{2} \textbf{Query Preprocessing.} Given a query program $q$, like the first step \blackding{1}, the program-level BCSA method $Enc(q)$ to get its representation;

\noindent \blackding{3} \textbf{Retrieving.} We send $q$ to the similarity search system (e.g., FAISS~\cite{douze2024faiss} and Milvus~\cite{2021milvus, 2022manu}) built based on $R$ to retrieve the most Top-$k$ similar programs from $R$ by leveraging program representations with function $F$. The search procedure has many indexes~\cite{2021milvus} such as \texttt{FLAT}, \texttt{HNSW}, \texttt{IVF\_FLAT}, and so forth. In this study, to accurately evaluate the performance of program-level BCSA methods, we use the \texttt{FLAT} index by default through brute force search.


\subsection{Motivation}\label{sec:motivation}

In this section, we introduce the motivation behind the design of \keenhash{} for \textbf{large-scale program-level BCSA}. Binary diffing~\cite{duan2020deepbindiff} is a widely used method to identify differences between two binaries, enabling various analyses. Where, function-level binary diffing (i.e., similar function matching)~\cite{ghidra, binaryai, diaphora} is popular since functions represent sufficiently detailed information about the decomposition of the program functionality. Therefore, using the results of (similar) function matching proportion as a measure can effectively capture the structural similarity between binaries. However, the approach introduced in Sec. \ref{sec:introduction} costs a $O(nm^3)$ time complexity for $n$ similarity evaluations. Even with heuristic strategies, scalability remains unsatisfiable on large-scale BCSA scenarios (e.g., BinDiffMatch takes 56 days to evaluate 5.3 billion similarities. See Sec. \ref{sec:eval:functionmatching}). Nevertheless, considering that function matching is essentially a form of classification, grouping similar functions into the same class which is dynamically expanding. Thus, as long as a precise and comprehensive classifier is used to classify functions, the matching is equivalent to and transformed into classification. In addition, by encoding classified results as positions (e.g., Feature Hashing~\cite{weinberger2009feature}) in a vector, $n$ similarity comparisons only require $O(n) \ll O(nm^3)$ time complexity where the dimension (length) of the vector is fixed. With this property, we can hash any binary, with a varying number of functions, to a compact and fixed-length vector, with hardware acceleration, to support fast comparison in large-scale scenarios. We propose \keenhash{}-stru based on this insight (Sec. \ref{sec:structurebasedembeddinggeneration}). Moreover, considering that massive code reuse (a widespread practice) can misleadingly increase the similarity between two non-same-class binaries (Sec. \ref{sec:introduction} and \ref{sec:semanticsbasedembeddinggeneration}), we introduce \keenhash{}-sem from the perspective of program semantics in Sec. \ref{sec:semanticsbasedembeddinggeneration} to mitigate this issue in large-scale BCSA.

\subsection{Assumption of Processed Binary}\label{sec:assumption}

In this section, we outline the assumption of the processed binary for \keenhash{}. Binary packing~\cite{dambra2023decoding} is the technique for compressing original binaries to reduce their size and obfuscate their contents.
%
%
Packed binaries are decompressed in memory during runtime before the original content in binaries is executed, and are difficult to accurately decompress statically~\cite{mantovani2020prevalence}. Directly using decompilers, like Ghidra~\cite{ghidra} or IDA Pro~\cite{idapro}, may not identify and analyze all functions within packed binaries.
\keenhash{} hashes programs based on decompiled results through Ghidra (see Sec. \ref{sec:functionextraction}) which may be affected by packing techniques.
Additionally, how to unpack any packed binary goes beyond the scope of our paper. Thus, to avoid biases, in this study, we divide the range of binaries that \keenhash{} can process into programs without packing, or packed programs can be unpacked easily (i.e., UPX~\cite{upx}).
%

\begin{figure*}[t]
    \centering
    \includegraphics[width=1.0\textwidth]{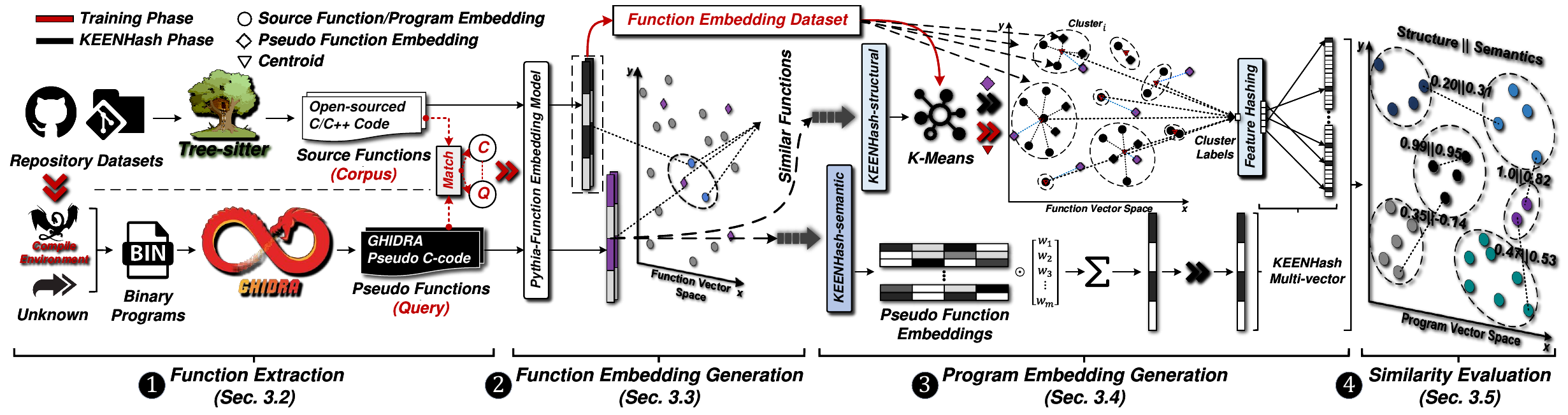}
    \caption{Workflow of \keenhash{}.}
    \label{fig:keenhashworkflow}
\end{figure*}

%% file: secs/methodology.tex
\section{Methodology}\label{sec:methodology}

In this section, we introduce the workflow of \keenhash{} for large-scale program-level BCSA.  

\subsection{Overview}\label{sec:overview}

As shown in Fig. \ref{fig:keenhashworkflow}, the workflow of \keenhash{} includes four phases: \blackding{1} function extraction (Sec. \ref{sec:functionextraction}), \blackding{2} function embedding generation (Sec. \ref{sec:functionembeddinggeneration}), \blackding{3} program embedding generation (Sec. \ref{sec:programembeddinggeneration}), and \blackding{4} similarity evaluation (Sec. \ref{sec:similarityevaluation}). Specifically, to hash a given binary to a multi-vector for similarity analysis, \blackding{1} \keenhash{} extracts the C-like pseudocode functions in the binary through Ghidra~\cite{ghidra}. Then, \blackding{2} \keenhash{} leverages a large language model (LLM) to generate the embeddings of these pseudo functions. Next, after obtaining the pseudo function embeddings, \blackding{3} \keenhash{} leverages the function matching-based structural feature (\keenhash{}-stru) as well as the function-intrinsic semantic feature (\keenhash{}-sem) to transform pseudo function embeddings into compact and fixed-length structural and semantic program embeddings, respectively, forming a multi-vector. Where the second is proposed by considering huge cose reuse cases (see Sec. \ref{sec:eval:programclonesearch}). Eventually, \blackding{4} \keenhash{} leverages the corresponding and respective similarity evaluation metrics for the generated structural and semantic embedding to compare the similarities of the given binary with other program embeddings, to support various large-scale program-level BCSA tasks.

\subsection{Function Extraction}\label{sec:functionextraction}

The initial step of \keenhash{} involves extracting C-like pseudocode functions (pseudo functions) from binaries where these pseudo functions are translated from corresponding binary functions in the binaries through Ghidra~\cite{ghidra}. The C-like pseudocode, rather than others (e.g., assembly code, byte code, and so forth)~\cite{wang2022jtrans, wang2024clap, ding2019asm2vec}, shields the details of assembly instructions from various architectures (i.e., unify to the same code format) and is close to the code in high-level languages such as C/C++ (i.e., one model can easily encode both source and pseudo functions in the same tokenization). Moreover, to train the subsequent LLM (function embedding model) for generating function embeddings, source functions from open-sourced C/C++ projects are extracted, with their pseudo functions compiled through various compile environments. Massive C/C++ source functions are also extracted for \keenhash{}-stru (see Sec. \ref{sec:structurebasedembeddinggeneration}) when generating program embeddings. Contrary to the intuitive imagination of \keenhash{} only hashing binaries, we do not adopt a binary-to-binary training method to LLM like previous works~\cite{wang2022jtrans, ding2019asm2vec} (i.e., with only pseudo/binary functions), but instead use a \textbf{\emph{source-to-binary}} approach (i.e., with both source and pseudo functions). The reason is three-fold:  (1) the language of C-like pseudocode is close to the languages in source code from C/C++ projects; (2) a source function acts as an anchor to many corresponding binary functions compiled from different compile environments; and (3) it is essential to support generating massive source function embeddings instead of pseudo ones (decompilation for binaries is time and resource-consuming) for \keenhash{}-stru (see Sec. \ref{sec:structurebasedembeddinggeneration}). Additionally, we only choose C/C++ open-sourced projects for fast construction of the automatic compilation pipeline inspired by jTrans~\cite{wang2022jtrans}, due to compilation compatibility reasons. The specific function extraction process is listed as follows:

\subsubsection{Training Phase}\label{sec:1trainingphase} 
\keenhash{} performs function extraction and ground-truth matching.

\noindent $\bullet$ \textbf{Source Function Extraction.} The process extracts C/C++ source functions from open-sourced C/C++ projects, such as GitHub repositories~\cite{github}. For each project, we collect all the C/C++ source code files across all versions through \code{git tags}. All source code files are deduplicated with the sha256 hash values. We also leverage tree-sitter~\cite{treesitter} to parse these files and extract all unique source functions (through the sha256 hash values of their content without comments and whitespaces) with line numbers in files from them. The mapping relationships among projects, project versions, source code files, line numbers, and source functions are preserved during the extraction.

\noindent $\bullet$ \textbf{Pseudo Function Extraction.} For a given binary, we leverage Ghidra~\cite{ghidra} to decompile it and extract all its binary functions that are translated to the language of C-like pseudocode (i.e., pseudo functions). Moreover, the relative virtual addresses (\textit{rva}) to the binary functions are also extracted.  

\noindent $\bullet$ \textbf{Source and Pseudo Function Matching.} Given a C/C++ project and corresponding compiled (unstripped) binaries through one specific compile environment (e.g., <GCC v13.2, O3, x86, 64-bit>), the matching process matches the source functions with their pseudo ones. A matched pair of the source and pseudo functions is an invertible mapping of the source one and its compiled binary one. Specifically, we compile the project and generate the debugging information (DWARF~\cite{dwarf}). Then, we perform pseudo function extraction to extract the mapping between pseudo functions and their \textit{rva}. Meanwhile, we parse the debugging information to extract the mapping between \textit{rva} and corresponding source files with line numbers. After further performing the source function extraction to the project and getting the third mapping, we can merge these mappings to get the $1$-to-$n$ matching from the source functions to the pseudo ones. Moreover, we use sha256 hashed values to deduplicate pseudo functions for each source. By performing matching and deduplication on extensive C/C++ projects and binaries, we get a \textit{Corpus} dataset \textit{C} and a \textit{Query} dataset \textit{Q} where they contain matched and unique source functions, and matched pseudo functions, respectively. The matching between \textit{C} and \textit{Q} is a $1$-to-$n$ mapping across binaries in one project, in different projects, and through various compile environments. \textit{C} and \textit{Q} are further used for training the subsequent function embedding model.


\subsubsection{Hashing Phase}\label{sec:1hashingphase}

Only binaries are processed to extract pseudo functions. Specifically, given a binary, we leverage Ghidra to decompile it and extract its pseudo functions, preparing to generate their function embeddings.

\subsection{Function Embedding Generation}\label{sec:functionembeddinggeneration}

The foundation of \keenhash{} is the function embedding model for generating function embeddings to both source and pseudo functions (Sec. \ref{sec:functionextraction}) for subsequent function-aware program embedding generation.  
The objective of the model is to generate function representations (embeddings) such that similar source and pseudo functions are gathered naturally in the vector (embedding) space. Conversely, dissimilar functions remain distanced from each other. To place the representations of both kinds of functions in the same space, we train our model based on the pairs of matched source and pseudo ones extracted from Sec. \ref{sec:1trainingphase}. While the grammar of source and pseudo code is similar~\cite{ghidra}, significant differences can still exist in the code due to various language features (e.g., function inlining~\cite{jia20231}), compile environments (e.g., optimization level~\cite{ding2019asm2vec}), and decompilation (e.g., accessing data members and functions~\cite{ghidra}), resulting in different formats. Recent work~\cite{nijkamp2022codegen, feng2020codebert} shows that existing LLMs, trained on code in various languages, can provide the capacity to understand and discriminate the intricate details and similarities of code syntax and semantics across different formats~\cite{allal2023santacoder, li2023starcoder, zeng2022extensive}. Therefore, we leverage LLM to overcome this issue and generate function embeddings. Instead of training a model from scratch, we use a pre-trained one for the transfer of knowledge~\cite{nijkamp2022codegen, guo2022unixcoder} and further fine-tune it on the pairs of matched functions through contrastive learning~\cite{radford2021learning, zhai2023sigmoid}. Through this approach, we enable the LLM to draw similar functions closer together while pushing dissimilar ones farther apart. Specifically, we leverage \pythia{} (contain 410M parameters)~\cite{pythia410m, biderman2023pythia}, a transformer-based language model widely adopted by the research community, as the initialized base model for further fine-tuning. Furthermore, we highlight that our function embedding model differs from existing state-of-the-art ones, such as jTrans~\cite{wang2022jtrans} and CLAP~\cite{wang2024clap}, in its ability to map both source and pseudo (binary) functions into the same space, while theirs only support binary ones. Our model is also more effective than theirs, focusing on binary function embeddings and correspondingly generated program embeddings (see Appendix~\ref{app:discussion}).


\subsubsection{Training Phase}\label{sec:2trainingphase}

In the training phase, we further fine-tune \pythia{} in a supervised manner on the pairs of matched functions through contrastive learning. Contrastive learning~\cite{chen2020simple} is a technique, engaging in-batch negative samples, for a model to learn an embedding space where similar sample pairs stay close to each other while dissimilar ones are far apart, leading to better performance on discriminating functions~\cite{shi2023cocosoda}. In particular, we leverage Contrastive Language-Image Pre-training (CLIP)~\cite{radford2021learning, zhai2023sigmoid} for fine-tuning our model due to the different code formats of source and pseudo functions. Specifically, \blackding{1} we first perform tokenization on all source and pseudo functions in the training dataset that converts them into sequences of tokens. \blackding{2} In the training epochs, batches are generated randomly and dynamically for more effective learning~\cite{zhai2023sigmoid}. Thus, to generate one batch, we randomly sample $N$ pairs of matched similar source and pseudo functions from \textit{C} and \textit{Q} with the near-deduplication procedure used in StarCoder~\cite{li2023starcoder} for more diverse training data, where functions between pairs are considered dissimilar. We pass the tokenized sequences of the $2N$ functions to \pythia{} and obtain their embeddings by extracting the $n$-dimensional output of the last hidden layer of the model where $n=1024$ (each dimension in \texttt{float32}, i.e., 4 bytes)~\cite{pythia410m}. Here, we denote $\mathbf{e}_{i}^s$ and $\mathbf{e}_{j}^p \in \mathbb{R}^n$ as the embeddings of the $i_{th}$ and $j_{th}$ source and pseudo functions, respectively. Furthermore, $\mathbf{e}_{i}^s$ and $\mathbf{e}_{j}^p$ are considered a match (i.e., positive pair) if $i$ is equal to $j$; otherwise, they are deemed unmatched (i.e., negative pair). \blackding{3} In each batch, CLIP evaluates the $N \times N$ cosine similarity matrix~\cite{zhai2023sigmoid} between all the pairs of functions based on their embeddings. \blackding{4} The training objective is to generate function embeddings in such a way that the similarity values of $N$ positive pairs are maximized, while the similarity values of the $N \times (N - 1)$ negative pairs are minimized. Therefore, we apply the softmax loss for language image pre-training~\cite{zhai2023sigmoid}, across source and pseudo functions, to the previously generated cosine similarity matrix. The specific loss function is defined as follows:

\begin{equation}
    \label{equ:lossfunction}
    \mathcal L = - \frac{1}{2N} \sum^{N-1}_{i=0} (\overbrace{\log \frac{e^{t \mathbf{x}_i \cdot \mathbf{y}_i}}{\sum^{N-1}_{j=0} e^{t \mathbf{x}_i \cdot \mathbf{y}_j}}}^{\text{source} \rightarrow \text{pseudo softmax}} + \overbrace{\log \frac{e^{t \mathbf{x}_i \cdot \mathbf{y}_i}}{\sum^{N-1}_{j=0} e^{t \mathbf{x}_j \cdot \mathbf{y}_i}}}^{\text{pseudo} \rightarrow \text{source softmax}})
\end{equation}

\noindent Where $\mathbf{x}_i = \mathbf{e}_{i}^s / \lVert\mathbf{e}_{i}^s\rVert_{2}$ and $\mathbf{y}_j = \mathbf{e}_{j}^p / \lVert\mathbf{e}_{j}^p\rVert_{2}$; $t$ is a freely learnable parameter for scaling logits~\cite{radford2021learning}.


\subsubsection{Hashing Phase}\label{sec:2hashingphase}

Given a bunch of pseudo functions extracted from a binary, \keenhash{} leverages the function embedding model to generate corresponding function embeddings, for subsequent program embedding generation.

\subsection{Program Embedding Generation}\label{sec:programembeddinggeneration}

Functions are self-contained code modules designed to perform specific tasks, and their combination constitutes the functionality and representation of a program. The core of \keenhash{} lies in integrating the functions in a binary to generate a compact and fixed-length embedding that represents the binary, for large-scale BCSA. Therefore, we approach it from two perspectives, respectively: \blackding{1} program structure (i.e., the function matching between binaries) and \blackding{2} program semantics (i.e., amplification of unique feature semantics between binaries when comparison). As mentioned in Sec. \ref{sec:motivation}, for \blackding{1}, the extent to which functions match between two binaries can serve as a metric for assessing their similarity. However, the time complexity of direct matching is prohibitively high (Sec. \ref{sec:motivation}). Thus, we transform the function matching problem into a classification one, to achieve matching at a much lower time complexity. As for \blackding{2}, due to the widespread practice of massive code reuse~\cite{luo2023vulhawk, jiang2023third, jiang2024binaryai, woo2021centris}, it often leads to the structure (e.g., function matching, call graph, and so forth) of two binaries appearing very similar. This results in difficulty in distinguishing binaries (in large-scale scenarios) in the same or different classes but with similar structures (see example in Sec. \ref{sec:eval:programclonesearch}) since the significance of the unique feature parts is indirectly weakened. Therefore, we explore the integration of function embeddings by capturing the semantic differences and maximizing the unique features among functions to effectively reflect the overall program semantics. We denote the first method as \keenhash{}-stru (Sec. \ref{sec:structurebasedembeddinggeneration}) and the second as \keenhash{}-sem (Sec. \ref{sec:semanticsbasedembeddinggeneration}). Together, their respective program embeddings combine into multi-vectors and can be selectively utilized based on specific conditions. Moreover, we highlight that both methods should be in an unsupervised manner due to the prohibitive costs of crafting large-scale and diverse labeled training datasets of binaries for real-world scenarios and generalization~\cite{benoit2023scalable, duan2020deepbindiff}.

\subsubsection{Structure-based Embedding Generation}\label{sec:structurebasedembeddinggeneration}

The insight of \keenhash{}-stru is performing function classification for function matching based on the pseudo functions to the given binary. Therefore, it is crucial to find a classifier that can efficiently classify a function such that similar functions are in the same class and dissimilar ones are separated. However, it is challenging to craft a high-quality training dataset for training a multi-classifier in a supervised manner due to the two aspects of labeling functions and determining the number of labels. To overcome this issue, we regard the classification as the $1$-NN search~\cite{douze2024faiss} by our function embedding model. Our model unifies source and pseudo functions within the same vector space, enabling us to perform the $1$-NN search on a vast training dataset crafted based on source functions to classify pseudo ones. The extraction of source functions is achieved from open-sourced C/C++ projects, eliminating the need for reverse engineering of binaries. This approach allows for the easy expansion of the $1$-NN training dataset, encompassing a wide variety of function semantics. Additionally, the model is capable of grouping similar functions together, while distinctly separating dissimilar ones, resulting in the massive collection of source functions forming clustered distributions directly. Therefore, by employing the suitable clustering algorithm, we can automatically extract function classes and label functions for the $1$-NN dataset. Moreover, to avoid the high time cost of $1$-NN search, centroid-based clustering is preferred where each cluster is represented by a centroid instead of all inner-cluster data points.

\noindent \textbf{Clustering.} Specifically, we first perform source function extraction (see \ref{sec:1trainingphase}) on massive projects to extract a large collection of distinct C/C++ source functions. Next, our trained function embedding model is applied to these functions to generate function embeddings. These embeddings are used as the training dataset for the subsequent centroid-based clustering. Here, we utilize K-Means~\cite{arthur2007k}, an effective unsupervised algorithm scalable on large-scale datasets, to perform clustering. Formally, the K-Means clustering algorithm partitions the training dataset of the source functions into $n$ clusters $S = \{S_0, S_1, ..., S_{n - 1}\}$ by maximizing the cosine similarity between functions in the same cluster. $c_i \in C$ is the centroid and representation for the cluster of $S_i$, labeled with $i$. $C$ is the set of all centroids, in the size of $n \ll \text{the number of source functions}$ that serves as the training dataset of $1$-NN. While training the K-Means model is expensive, it is a one-time and offline process in a period that does not impact the efficiency of \keenhash{}-stru. In addition, the cluster size of $n$ to K-Means is a critical hyperparameter that can affect the performance of the subsequent classification task. Recall that our objective is to transform the matching problem into the classification one, and therefore the performance of the matching task is the evaluation metric for finding the most effective $n$ to train the K-Means model. To this end, we set $n$ to $2^k$ where $k \in \mathbb{N}^+$ to appropriately reduce its search range. We systematically study a suitable $n$ in the function matching task between binaries in the same class in Sec. \ref{sec:eval:functionmatching}.

\noindent \textbf{Generation.} After obtaining the $n$ centroids, we take them as the training dataset of $1$-NN and perform the $1$-NN search to all pseudo functions of the binary for generating the program embedding. Formally, we denote the collection of pseudo functions with $q$ numbers as $P = \{P_0, P_1, ..., P_{q-1}\}$ where $P_i$ is the $i_{th}$ pseudo function. Then after the search, we get the corresponding collection of retrieved Top-$1$ most similar centroids $R = \{c^0_i, c^1_j, ..., c^{q-1}_k\}$ with cosine similarity where $c^i_j$ represents that $P_i$ retrieve the $c_j \in C$ centroid. Therefore, taking the labels $L = \{i, j, ..., k\}$ of $R$, the binary is transformed into a collection of labels, and for any two binaries, their respective pseudo functions with the same labels are considered matches. Additionally, to further enhance the efficiency of similarity evaluation with bitwise operations, we attempt to transform $L$ of the binary into a fixed-length bit-vector $\mathbf{v}$ in dimension $n$: $\mathbf{v} = [v_0, v_1, ..., v_{n-1}]$ where $v_i = 1$ if $i \in L$; otherwise, $v_i = 0$. This transformation captures the presence of each centroid in the result of the $1$-NN search, and multiple identical labels are consolidated as a single one since the pseudo functions are classified into the same class, potentially presenting similar semantics. Moreover, each element in $\mathbf{v}$ is represented as a single bit to minimize space requirements.

However, the dimension $n$ of $\mathbf{v}$ can be very high (e.g., $n=2^{22}$ in 512KB), and the number of pseudo functions in the binary can be extremely smaller. Directly using the dimension $n$ can consume a substantial amount of space and is computationally prohibitive~\cite{hu2013mutantx}. Furthermore, the size of $n$ for K-Means may not be reducible due to the consideration regarding the performance of function matching (Sec. \ref{sec:eval:functionmatching}). To address this issue, we employ Feature Hashing~\cite{weinberger2009feature}, which hashes the high-dimensional input vector $\mathbf{v} \in \{0, 1\}^n$ into a lower one $\{0, 1\}^m$ with the mapping function $\phi: \mathcal V \rightarrow \{0, 1\}^m$ where $\mathcal V$ is the domain of all possible $\mathbf{v}$. Since $m \ll n$, Feature Hashing reduces $\mathbf{v}$ to a more compact representation, allowing for significant savings in space and computational resources. Moreover, previous research demonstrates that Feature Hashing approximately preserves the original similarity (i.e., function matching results) between hashed vectors with a high probability~\cite{jang2011bitshred}, and the penalty incurred from using it only grows logarithmically with the number of samples compared~\cite{weinberger2009feature}.

In particular, according to $\mathbf{v}$, we leverage a uniform hash function $H: \mathcal L \rightarrow [0, m)$ to hash a label $i \in \mathcal L \land v_i = 1$ ($\mathcal L$ is the domain of all possible labels, i.e., all clustered centroids) to the new position $j \in [0, m)$ of hashed bit-vector $\mathbf{v}' = [v'_0, v'_1, ..., v'_{m-1}]$ in dimension $m$. Furthermore, a sign hash function $\zeta: \mathcal L \rightarrow \{-1, +1\}$ is applied to the label $i$ to get its signed value for leading to an unbiased estimate~\cite{weinberger2009feature}. In case of collision where multiple labels map to the same position $j$, the sum of their signed values, followed with an indicator function $\mathbf{1}_{x \neq 0}(x)$ for preserving bitwise operations and space requirements, is taken as the value for $v'_j \in \{0, 1\}$. Formally, for a given $\mathbf{v}$, the $\phi$ to get $\mathbf{v}'$ is defined as follows:

\begin{equation}
    v'_j = \phi_j(\mathbf{v}) = \mathbf{1}_{x \neq 0}(0 + \sum_{k \in \{i | H(i) = j \land v_i = 1 \}} \zeta(k))
\end{equation}

\noindent where $\mathbf{1}_{x \neq 0}(x) = 1$ if $x \neq 0$; otherwise, $\mathbf{1}_{x \neq 0}(x) = 0$.

In addition, the selection of the hashed length $m$ is critical since it balances the efficiency and effectiveness, as well as space requirements. Based on our experience, we aim to select a larger $m$ as much as possible. Specifically, we set $m=2^{16}$ (e.g., 512KB to 8KB), which is a reasonable upper bound of space size and applies to the vector database Milvus~\cite{2021milvus}, supporting at most $2^{18}$ in bits. We also perform a discussion on the selection of $m$ in Appendix~\ref{app:hashvaluesize}.
%

\subsubsection{Semantics-based Embedding Generation}\label{sec:semanticsbasedembeddinggeneration}

\keenhash{}-stru compares binaries through the structural features of function matching. However, the semantic differences between functions are normalized to only two states: matched or unmatched, i.e., any function is considered equally significant. Thus, structurally similar (i.e., massive code reuse~\cite{jiang2023third}) binaries in the same or different classes (e.g., across compile environments. See the previously mentioned example in Sec. \ref{sec:eval:programclonesearch}) may affect \keenhash{}-stru to some extent of distinguishing them in large-scale BCSA (Sec. \ref{sec:eval:programclonesearch}), due to lacking the capability of simultaneously capturing the semantic differences and maximizing the unique features. This problem is also faced by other structure-based methods with even more negative impact (Sec. \ref{sec:eval:programclonesearch}). To mitigate this issue, we introduce \keenhash{}-sem which integrates function semantics based on significance to derive the program semantics.



\noindent \textbf{Generation.} A potential way for generating the semantic embedding to a binary involves averaging directly the pseudo function embeddings (Mean Pooling)~\cite{arora2017simple}. However, this strategy ignores the quantity of information for each function such as the size of a function~\cite{yang2022modx}. Inspired by the success of TF-IDF~\cite{wu2008interpreting} and SIF~\cite{arora2017simple} techniques which model sentence semantics based on the weighted average of word embeddings to express the most significant words, we propose a similar strategy to derive program semantics through function ones. Notably, the feature functions are exclusively included in the same-class binaries and not present across any non-same-class ones, i.e., the unique features. Thus, we utilize the intrinsic information of the function as weights to assess its significance and thereby, maximize the semantics of feature functions for amplifying the unique similarities (the feature functions between same-class binaries) or differences (the respective feature functions between non-same-class ones) while offsetting the ones of the reused for stronger distinguishing ability. Specifically, we determine the weights for a pseudo function by modeling based on two effective factors of its lines of code (LoC) and the number of strings (NoS). Guided by heuristics, we posit that a pseudo function's importance in a binary increases with its LoC and NoS since a function with a higher LoC likely handles more complex logic, making it crucial within the overall program and a larger NoS may indicate extensive functionality in processing user inputs, displaying data, or executing other tasks that heavily involve string operations. Therefore, for a pseudo function $P_i$ with its $\text{LoC}_i$ and $\text{NoS}_i$, its weight $w_i$ is specified as follows:



\begin{equation}
    w_i = f_1(\text{LoC}_i, \alpha) + f_2(\text{NoS}_i, \beta)
\end{equation}

\noindent where $f_1$ and $f_2$ are designed to compute the partial weights based on $\text{LoC}_i$ and $\text{NoS}_i$, respectively. $\alpha$ and $\beta$ are hyperparameters that adjust and scale the influence of $\text{LoC}_i$ and $\text{NoS}_i$ on the overall weight of $P_i$. Typically, $f_1$ and $f_2$ are determined empirically and experimentally. We evaluate the performance of program clone search on the IoT (malicious) and BinaryCorp (benign) repository datasets (Sec. \ref{sec:dataset}), without their query parts, to adjust and find an optimal configuration for them:

\begin{equation}
f_1(\text{LoC}_i, \alpha) = \frac{(\text{LoC}_i)^{\alpha_1}}{\alpha_2}\text{,\quad}f_2(\text{NoS}_i, \beta) = \frac{(\text{NoS}_i)^{\beta_1}}{\beta_2} + 1
\end{equation}

\noindent where $\alpha_1, \alpha_2, \beta_1, \beta_2 = 0.4, 5, 0.45, 1$ (obtained through grid search~\cite{liashchynskyi2019grid}). For a binary with its pseudo functions $P = \{P_0, P_1, ..., P_{q-1}\}$ and corresponding function embeddings $E = \{\mathbf{e}^p_0, \mathbf{e}^p_1, ..., \mathbf{e}^p_{q-1}\}$, its program embedding is formulated as:

\begin{equation}
    \mathbf{v} = \frac{1}{q} \sum_{i=0}^{q-1} w_i \frac{\mathbf{e}^p_i}{\lVert\mathbf{e}_{i}^p\rVert_{2}}
\end{equation}

\noindent With this approach, \keenhash{}-sem can maximize the feature function semantics within the same-class binaries through the weights to effectively distinguish them from other non-same-class ones but with massive code reuse (see Sec. \ref{sec:eval:programclonesearch}).
Moreover, we have experimented with other features to set weights, such as \revision{\whiteding{1}} the vertex centrality in call graph \revision{and \whiteding{2} system API call usages (e.g., the count of API calls in a function)}.
However, the experimental results show that using them for setting weights is \revision{good but} less effective than LoC and\revision{/or} NoS \revision{(e.g., \whiteding{2})}, \revision{or is} even less than Mean Pooling \revision{(e.g., \whiteding{1})}. \revision{In this study, we do not delve into these features.}
%



\subsection{Similarity Evaluation}\label{sec:similarityevaluation}

Here, we present the similarity evaluation metric used by \keenhash{} for comparing two binaries.

\noindent \textbf{\keenhash{}-stru.} The program embedding produced by \keenhash{}-stru is a bit-vector, with each element indicating the classification-based function matching. We leverage Jaccard similarity~\cite{jang2011bitshred} to evaluate similarities due to its proportional measure property (see Appendix~\ref{app:sim}).


\noindent \textbf{\keenhash{}-sem.} For the float vector of \keenhash{}-sem, we leverage cosine similarity, which is consistent with the comparison between function embeddings~\cite{arora2017simple}. 

%% file: secs/evaluation.tex
\section{Evaluation}\label{sec:evaluation}

In this section, we attempt to investigate \keenhash{} on its performance of large-scale program-level BCSA (see Sec. \ref{sec:problemdefinition}) by answering the following research questions:

\noindent $\bullet$ \textbf{RQ1:} How \textit{effective} including \textit{scalable} is \keenhash{} in the task of function matching for large-scale BCSA scenarios?

\noindent $\bullet$ \textbf{RQ2:} How does \keenhash{} perform on program clone search with various respective large-scale repositories?

\noindent $\bullet$ \textbf{RQ3:} Is \keenhash{} effective against code obfuscation on large-scale program clone search?

\noindent $\bullet$ \textbf{RQ4:} Is \keenhash{} effective on program clone search with \textit{larger-scale} repository?

\noindent $\bullet$ \textbf{RQ5:} How does \keenhash{} perform on malware detection from the large-scale BCSA and clone search perspective?


\subsection{Dataset}\label{sec:dataset}

In this section, we briefly introduce our datasets and the details of them can be found in Appendix~\ref{app:datasetindetail}.

\noindent \textbf{Training Dataset.} Two training datasets are included for \blackding{1} the function embedding model and \blackding{2} the K-Means model, respectively. For \blackding{1}, we collect (and build) open-sourced C/C++ projects (along with corresponding collected binaries across various architectures if possible) through ArchLinux official repositories (AOR)~\cite{archlinux}, Arch User Repository (AUR)~\cite{archuserrepository}, and Linux Community~\cite{gnulinuxcommunity}, ultimately amassing around 910K projects. The source functions (with matched pseudo ones) related to the evaluation of RQ1 and the effectiveness of our function embedding model (see Appendix~\ref{app:discussion}) are excluded, preventing data leakage. Eventually, we obtain 40.51M matched function pairs with an average of 556 tokens per function. As for \blackding{2}, we follow previous studies~\cite{tang2022towards, wu2023ossfp} to collect a large number of diverse open-source C/C++ projects by crawling from Github~\cite{github} and GNU/Linux community~\cite{gnulinuxcommunity}. In total, 11,013 projects, including malicious ones (e.g., gh0st RAT malware~\cite{gh0st}), are obtained, containing 56M unique C/C++ source functions. Such a substantial source function dataset is essential for the generalization of \keenhash{}-stru.

\noindent \textbf{Test Dataset.} The test dataset is used to evaluate the performance of \keenhash{} on function matching (Sec. \ref{sec:eval:functionmatching}). Specifically, we use the binary diffing dataset in DeepBinDiff~\cite{duan2020deepbindiff}. In total, there are 2,098 binaries across various versions and optimization levels. The function matching ground truth is obtained through the Function Extraction (Sec. \ref{sec:functionextraction}).

\noindent \textbf{Repository and Query Dataset.} To evaluate \keenhash{} on BCSA (Sec. \ref{sec:problemdefinition}), we collect 5 datasets:

\noindent $\blacktriangleright$ \textbf{IoT.} We collect recent 37,657 nonpacked C/C++ (detected with DIE~\cite{die}) IoT malware samples from MalwareBazaar~\cite{malwarebazaar} across 21 malware families. We randomly divide the dataset into repository and query datasets in a 9:1 ratio and each family has at least two samples in the query.

\noindent $\blacktriangleright$ \textbf{BinaryCorp (BC).} The dataset~\cite{wang2022jtrans} is crafted based on AOR and AUR across 5 optimization levels. There are 9,819 source code and 45,593 distinct C/C++ binaries with 9,498 sample families. We randomly divide the dataset into repository and query in a 7.5:2.5 ratio and each family has at least one in the query. All binaries are stripped. Notably, it contains massive code reuse (Sec. \ref{sec:eval:programclonesearch}).

\noindent $\blacktriangleright$ \textbf{BinKit (BK).} BinKit~\cite{kim2022revisiting} dataset is crafted from 51 GNU software packages with 235 unique C/C++ source code (i.e., sample families). It is diverse along different optimization levels, compilers, architectures, and obfuscations. Like BinaryCorp, it also contains massive code reuse (Sec. \ref{sec:eval:programclonesearch}).

    $\bullet$ \textbf{Normal (BinKit$_N$/BK$_N$):} The normal one is compiled with 288 different compile environments for 67,680 binaries to 51 packages. It covers 8 architectures (arm, x86, mips, and mipseb, each available in 32 and 64 bits), 9 compilers (5 versions of GCC and 4 versions of Clang), and 4 optimization levels (O0, O1, O2, and O3);

    $\bullet$ \textbf{Obfuscation (BinKit$_O$/BK$_O$):} The obfuscation one is compiled with 4 obfuscation options, instruction substitution (SUB), bogus control flow (BCF), control flow flattening (FLA), and all combined (ALL), through Obfuscator-LLVM~\cite{ieeespro2015-JunodRWM}. The same architectures and 5 optimization levels (extra Os) are also covered, and 37,600 binaries are generated.

The BK$_N$ dataset is divided randomly into repository and query in a 9:1 ratio. 10\% of the samples are randomly selected from the BK$_O$ dataset as the query to maintain experimental consistency in Sec. \ref{sec:eval:againstobfuscation} and \ref{sec:eval:all}. All binaries are stripped.

\noindent $\blacktriangleright$ \textbf{MLWMC (MC).} MLWMC~\cite{dambra2023decoding} is a recent real-world PE 32 malware dataset. It contains 67,000 malware samples across 670 malware families. We consider the 49,820 nonpacked C/C++ samples, belonging to a total of 615 malware families where each family contains at least 20 samples. Moreover, we divide the dataset in the same way as IoT.

\subsection{Experiment Setup}\label{sec:experimentsetup}

\noindent \textbf{Function Embedding Model.} The maximum length of the embedding model is 2048, the training epoch is 196, the batch size (i.e., $N$) is 512, and the learning ratio is 0.001.

\noindent \textbf{K-Means Model.} The cluster size is set to $2^k$ where $k \in [16, 22]$ with the iteration of 30. Furthermore, we use FAISS-GPU~\cite{johnson2019billion} to train K-Means models. Thus, the maximum cluster size is limited to $2^{22}$ due to the VRAM constraint. The clustering takes at most 20 hours for the size of $2^{22}$.

\noindent \textbf{\keenhash{}.} The parameter settings are shown in Sec. \ref{sec:programembeddinggeneration}.

\noindent \textbf{Baseline.} For RQ1, we use popular open-sourced Diaphora~\cite{diaphora}\revision{, the most recent state-of-the-art academic SigmaDiff~\cite{gao2024sigmadiff},} and commercial BinDiffMatch~\cite{binaryai} tools for \emph{function matching} between binaries. Diaphora uses function hash values and calling relationships. \revision{SigmaDiff and} BinDiffMatch employs function embeddings, generated through deep/machine learning (DL/ML), and a call graph-based heuristic strategy to find the most similar functions. \revision{There are other relevant tools such as DeepBinDiff~\cite{duan2020deepbindiff} for direct binary diffing, and Asm2Vec~\cite{ding2019asm2vec} and PalmTree~\cite{li2021palmtree} for function embedding generation. In this study, we do not include them since SigmaDiff is an upgrade of DeepBinDiff~\cite{gao2024sigmadiff, duan2020deepbindiff}, and both SigmaDiff and BinDiffMatch have already covered function embedding generation for function matching, better than Asm2Vec and PalmTree.}


For RQ2 to RQ5, we include 4 state-of-the-art structure-based methods as our baselines: SSDEEP~\cite{kornblum2006identifying}, TLSH~\cite{oliver2013tlsh}, Vhash~\cite{virustotal}, and \psso~\cite{benoit2023scalable}. SSDEEP and TLSH are two fuzzy hash algorithms (on whole files) that are widely used in binary similarity evaluation. Vhash, a widely-recognized security BCSA method, is an in-house similarity clustering algorithm, based on a simple structural feature hash. It empowers VirusTotal to find similar files and perform threat intelligence. However, VirusTotal does not provide the information of similarity space. Thus, we define the Vhash in the Hamming space due to the best performance in experiments. \psso{} is a spectral-based method. It represents a binary by calculating the spectrum of its call graph and the edge counts in the control flow graphs (CFGs) of functions. \revision{In this study, we include SSDEEP and TLSH (belonging to fuzzy hashing families) due to their popularity in the industry for large-scale program-level BCSA~\cite{virustotal, tlsh, malwarebazaar}, though previous works~\cite{haq2021survey, wang2024we} claim that they may have poor performance. For comprehensive comparison, therefore, we incorporate the most recent methods from both industry and academia (i.e., Vhash and \psso).}
We also leverage \revision{4} methods to compare with \keenhash{} for the ablation study purpose in RQ2 and RQ3: Mean Pooling (Sec. \ref{sec:semanticsbasedembeddinggeneration})\revision{, \text{KEENHash$_\text{w/oFH}^{16}$}, LoC, and NoS}. \revision{\text{KEENHash$_\text{w/oFH}^{16}$} is the variant of \keenhash{}-stru but with a cluster size of $2^{16}$ for K-Means and thereby without Feature Hashing, for assessing the necessity of the Feature Hashing module. LoC and NoS represent using either one of the features in \keenhash{}-sem, for demonstrating their respective contribution.}
Here, we do not include function matching (and DL/ML)-based BCSA methods for comparison, such as \revision{SigmaDiff~\cite{gao2024sigmadiff},} BinDiffMatch~\cite{binaryai}, and \revision{other function embedding methods~\cite{ding2019asm2vec, li2021palmtree, wang2024clap}}, since they do not support direct use in large-scale scenarios (Sec. \ref{sec:eval:functionmatching}). \revision{The effectiveness of our function embedding model and correspondingly generated program embeddings, compared with others, are discussed in Appendix~\ref{app:discussion}, as mentioned before.} We also take no string-based method in as it lacks robustness and is vulnerable to simple string-based attacks (Sec. \ref{sec:eval:programclonesearch}).



\noindent \textbf{Experiment Environment.} All the experiments are run on a Linux server running Ubuntu 20.04 with AMD EPYC 7K62 48-Core Processor, 1TB RAM, and 8 Nvidia A100 GPUs. The program clone search is implemented atop Milvus~\cite{2021milvus}.

\subsection{RQ1: Effectiveness and Scalability of Function Matching}\label{sec:eval:functionmatching}

\begin{figure}[t]
\begin{minipage}[t]{0.48\textwidth}
\vspace{0em}
\captionof{table}{The performance of function matching for K-Means of \keenhash{}-stru in different cluster size settings.}
\label{tab:functionmatching}
\centering
\scalebox{0.6}{\begin{tabular}{c|c|c|c}
\toprule
\textbf{Method}                                  & \textbf{Precision} & \textbf{Recall} & \textbf{F1-Score}  \\
\midrule
\multicolumn{1}{l|}{Diaphora~\cite{diaphora}}                    & 0.7121             & 0.5944 & 0.6480    \\
\multicolumn{1}{l|}{\revision{SigmaDiff}~\cite{gao2024sigmadiff}}        & \revision{0.8640}  & \revision{0.7827} & \revision{0.8213}    \\
\multicolumn{1}{l|}{BinDiffMatch~\cite{binaryai}} & 0.9652             & 0.8870 & 0.9244    \\
\midrule
\multicolumn{1}{l|}{K-Means ($n$=$2^{22}$)}        & 0.7573             & 0.7730 & 0.7651    \\
\multicolumn{1}{l|}{K-Means ($n$=$2^{21}$)}        & 0.7282             & 0.7714 & 0.7492    \\
\multicolumn{1}{l|}{K-Means ($n$=$2^{20}$)}        & 0.7023             & 0.7656 & 0.7326    \\
\multicolumn{1}{l|}{K-Means ($n$=$2^{19}$)}        & 0.6403             & 0.7542 & 0.6926    \\
\multicolumn{1}{l|}{K-Means ($n$=$2^{18}$)}        & 0.6113             & 0.7182 & 0.6605    \\
\multicolumn{1}{l|}{K-Means ($n$=$2^{17}$)}        & 0.5675             & 0.7016 & 0.6274    \\
\multicolumn{1}{l|}{K-Means ($n$=$2^{16}$)}        & 0.5295             & 0.7138 & 0.6080    \\
\bottomrule
\end{tabular}
}

\captionof{table}{\revision{Statistics of function matching time cost for K-Means ($n$=$2^{22}$), \revision{SigmaDiff}, and BinDiffMatch.}}
\label{tab:functionmatchingtimecost}
\centering
\scalebox{0.5}{\begin{tabular}{c|c|c|c}
\toprule
\revision{\textbf{Method}} & \revision{\textbf{Mean Cost (1 core)}} & \revision{\textbf{RQ1 Scenario (1 core)}} & \revision{\textbf{RQ4 Scenario (48 cores)}}  \\
\midrule
\multicolumn{1}{l|}{\revision{SigmaDiff}}    & \revision{0.25074 seconds}  & \revision{483 seconds} & \revision{323 days}    \\
\multicolumn{1}{l|}{\revision{BinDiffMatch}} & \revision{0.04314 seconds}  & \revision{83 seconds} & \revision{56 days}    \\
\multicolumn{1}{l|}{\revision{K-Means ($n$=$2^{22}$)}}      & \revision{0.00020 seconds}  & \revision{0.38574 seconds} & \revision{395.83 seconds}    \\
\bottomrule
\end{tabular}
}

\end{minipage}%
\quad
\begin{minipage}[t]{0.48\textwidth}
\centering
\vspace{1em}
\includegraphics[width=0.95\textwidth]{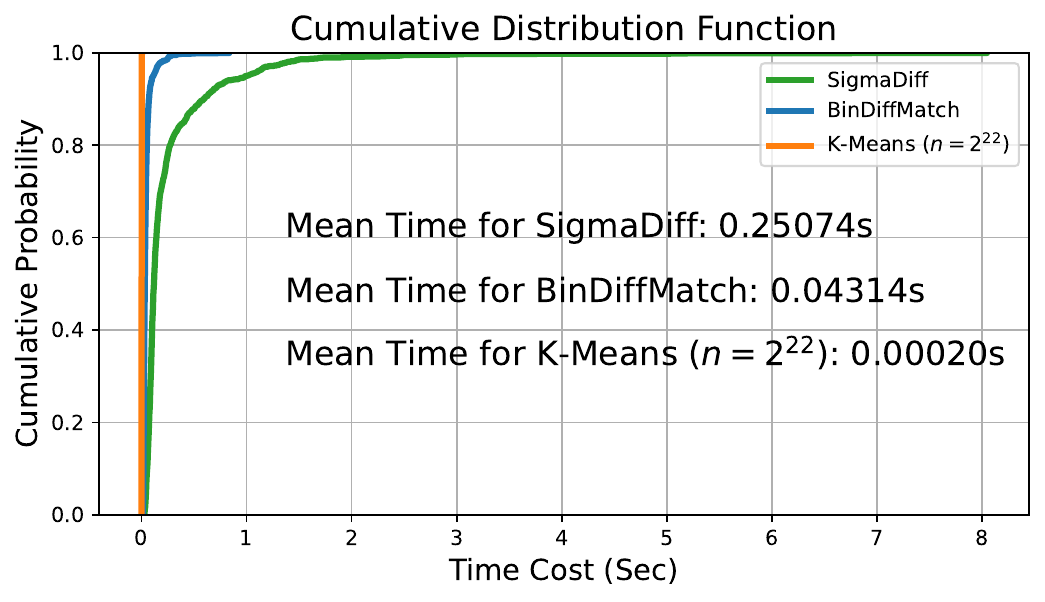}
\caption{The time cost of performing function matching for K-Means ($n$=$2^{22}$), \revision{SigmaDiff}, and BinDiffMatch with 1 core. K-Means ($n$=$2^{22}$) takes an average of 0.00020 seconds to complete one matching between two binaries, while \revision{SigmaDiff} and BinDiffMatch require \revision{0.25074} and 0.04314 seconds.}
\label{fig:timecostmatch}
\end{minipage} 
\end{figure}

\noindent \textbf{Motivation.} For \keenhash{}-stru, it is essential to choose an effective K-Means model to reflect the results of function matching accurately. Furthermore, we show that \keenhash{}-stru is scalable on \textit{large-scale} BCSA scenarios but existing state-of-the-art function matching methods are not.

\noindent \textbf{Approach.} We train multiple K-Means models by selecting various cluster sizes of $2^k$. These K-Means models are used for binary diffing to match same-class pseudo functions (compiled from the same source functions) between two same-class binaries. Same-class pseudo functions should be classified into the same clusters, whereas those from different classes should be assigned to distinct clusters. The methods used for comparison are Diaphora\revision{, SigmaDiff,} and BinDiffMatch. The compared pairs of binaries can be found here \cite{bindiffmatchgithub}, aligning with BinDiffMatch. In total, there are 1,926 pairs of binaries with 101,289 pairs of matched functions. The evaluation metrics include Precision (the ratio of corrected matches to all derived results), Recall (the ratio of corrected matches to all the ground-truth data), and F1-Score (the harmonic mean of Precision and Recall). We also measure the time cost for function matching to compare their scalability in large-scale scenarios.

\noindent \textbf{Result.} Table \ref{tab:functionmatching} shows that the performance of function matching with our K-Means models is enhanced as the cluster size increases. In particular, the K-Means ($n$=$2^{22}$) achieves the highest 0.7573, 0.7730, and 0.7651 in Precision, Recall, and F1-Score, significantly outperforming Diaphora. Compared to \revision{SigmaDiff and} BinDiffMatch, the performance of the K-Means is relatively poorer, but still able to maintain an effective capacity of \revision{93\% to SigmaDiff and} 83\% to BinDiffMatch in F1-Score (\textit{and it is the only one that supports large-scale scenarios}. See later). This is deemed reasonable as the K-Means models are unable to distinguish the similarities among classified functions and do not incorporate call graphs from binaries to provide essential diffing information for matching functions, resulting in a lower Recall. The classification nature can also result in multiple matched functions being classified into the same clusters, raising the rate of false positives (cartesian product) and lowering Precision. 

\noindent $\bullet$ \textbf{Scalability.} Additionally, we plot the cumulative distribution of the time cost for only function matching procedure of pairs of binaries in Fig. \ref{fig:timecostmatch} \revision{among SigmaDiff,} BinDiffMatch, and K-Means ($n$=$2^{22}$) with 1 core (Diaphora is out of the bound of 1 second). \revision{Table \ref{tab:functionmatchingtimecost} shows the statistics of time cost for them on function matching across mean time for one matching, RQ1 scenario total time cost, and RQ4 scenario (Sec. \ref{sec:eval:all}) total time cost.} The function embeddings, call graphs, and function classifications (K-Means) are generated offline, which is reasonable in large-scale scenarios. The results show that K-Means is on average \revision{1,254 times and} 215 times faster than \revision{SigmaDiff and} BinDiffMatch\revision{, respectively}. Particularly, in the large-scale \revision{RQ4} scenario of Sec. \ref{sec:eval:all} \revision{(see Table \ref{tab:functionmatchingtimecost})}, \revision{SigmaDiff and} BinDiffMatch will \revision{cost 323 and} 56 days to perform 5.3 billion similarity evaluations (\textit{a typical workload within one day}) between binaries with 48 processes (48 cores) running in parallel, which is unscalable and unavailable. Whereas \keenhashstructural{16} (Sec. \ref{sec:eval:programclonesearch}) based on K-Means ($n$=$2^{22}$) takes only 395.83 seconds. Therefore, we consider the K-Means ($n$=$2^{22}$) model to be effective, as it provides an effective function matching capability and is well-suited for scalability for program-level BCSA on large-scale scenarios. As for Diaphora\revision{, SigmaDiff,} and BinDiffMatch, they are unable to support such large-scale scenarios and are thereby excluded from the following RQs.

\mybox{\textbf{Answer 1:} \keenhash{}-stru can effectively (0.7651 in F1-Score) and scalably perform function matching in large-scale scenarios. While the state-of-the-art matching methods are not scalable, being at least 215 times slower. In a scenario with 5.3 billion evaluations, \keenhash{}-stru takes 395.83 seconds while they will cost at least 56 days.}

\subsection{RQ2: \keenhash{} on Program Clone Search}\label{sec:eval:programclonesearch}  

\noindent \textbf{Motivation.} In this RQ, we attempt to evaluate \keenhash{} performance on program clone search, reflecting its capacity in program-level BCSA on \revision{respective and different} large-scale scenarios.

\noindent \textbf{Approach.} We perform program clone search on 4 repository and query datasets (except BinKit$_O$), respectively. Here, we denote \keenhash{}-stru and sem as \keenhashstructural{16} (16 represents Feature Hashing size of $2^{16}$) and \keenhashsemantic{} ($f$ represents float vectors), respectively. The evaluation metrics include mAP@$k$ (Mean Average Precision at $k$) and mP@$k$ (Mean Precision at $k$)~\cite{carterette2011overview} where $k$ represents retrieving the most Top-$k$ similar results. mAP@$k$ is a metric that evaluates the mean (across all retrieving results) of the average of the Precision@$k_i$ ($k_i \in \{i | 1 \leq i \leq k \wedge r(i) = 1\}$ where $r(i) = 1$ represents that the $i_{th}$ retrieved sample is in the same class to the query sample; otherwise, $r(i) = 0$) to one retrieving. The higher the ranking of same-class results returned, the greater the mAP@$k$. However, mAP@$k$ cannot assess the proportion of same-class results returned. Therefore, we augment it with mP@$k$, which calculates the mean of the Precision@$k$ across all retrieving. Where Precision@$k_i$ presents the ratio of retrieved same-class samples to the retrieved Top-$k_i$ ones. Here, we fix $k = 100$ to evaluate the \keenhash{} BCSA capacity on multiple same-class binaries.

\begin{table}[t]
\begin{minipage}[t]{0.32\textwidth}
\vspace{0em}
\caption{Program clone search against string obfuscation. The string-based method \minhashstring{} is vulnerable to the simple attack.}

\centering
\label{tab:programclonesearchmirai}
\scalebox{0.5}{\begin{tabular}{ccc|cc}
\toprule

\multirow{2}{*}{\textbf{Method}}   & \multicolumn{2}{c|}{\textbf{mAP@100}} & \multicolumn{2}{c}{\textbf{mP@100}}     \\

\cmidrule{2-5}

                    & \multicolumn{1}{c}{{Mirai$_N$}} & \multicolumn{1}{c|}{{Mirai$_O$}} & \multicolumn{1}{c}{{Mirai$_N$}} & \multicolumn{1}{c}{{Mirai$_O$}} \\
\midrule

\multicolumn{1}{l|}{\minhashstring} & \textbf{1.0} & \textbf{0.1909} & \textbf{1.0} & \textbf{0.1040} \\

\midrule

\multicolumn{1}{l|}{\psso}                & 0.9977 & 0.9887 & 0.9960 & 0.9894 \\

\midrule

\multicolumn{1}{l|}{Vhash}               & 0.6442 & 0.6442 & 0.5350 & 0.5350 \\
\multicolumn{1}{l|}{TLSH}                & 0.8583 & 0.8472 & 0.8393 & 0.8398 \\
\multicolumn{1}{l|}{SSDEEP}              & 0.9901 & 0.9900 & 0.9900 & 0.9900 \\

\midrule

\multicolumn{1}{l|}{\keenhashstructural{16}} & 0.9997 & 0.9999 & 0.9996 & 0.9997 \\
\multicolumn{1}{l|}{\keenhashsemantic}       & 0.9937 & 0.9939 & 0.9883 & 0.9873 \\

\bottomrule
\end{tabular}}
\end{minipage}%
\quad
\begin{minipage}[t]{0.64\textwidth}
\vspace{0.2em}
\caption{The performance of program clone search of \keenhash{} on respective datasets of IoT, BinaryCorp, BinKit$_N$, and MLWMC.}
\centering
\label{tab:programclonesearch}
\scalebox{0.5}{\begin{tabular}{cccccc|ccccc}
\toprule

\multirow{2}{*}{\textbf{Method}}   & \multicolumn{5}{c|}{\textbf{mAP@100}} & \multicolumn{5}{c}{\textbf{mP@100}}     \\

\cmidrule{2-11}

                    & \multicolumn{1}{c}{{IoT}} & \multicolumn{1}{c}{{BC}} & \multicolumn{1}{c}{{BK$_N$}} & \multicolumn{1}{c}{{MC}} & \multicolumn{1}{c|}{{Avg.}} & \multicolumn{1}{c}{{IoT}} & \multicolumn{1}{c}{{BC}} & \multicolumn{1}{c}{{BK$_N$}} & \multicolumn{1}{c}{{MC}} & \multicolumn{1}{c}{{Avg.}} \\
\midrule

\multicolumn{1}{l|}{\psso}                & 0.9383 & 0.3212 & 0.6803 & 0.7164 & 0.6641 
& 0.8991 & 0.0141 & 0.3033 & 0.3666 & 0.3958 \\

\midrule

\multicolumn{1}{l|}{Vhash}               & 0.6789 & 0.5646 & 0.3379 & 0.7377 & 0.5798 
& 0.6455 & 0.0207 & 0.0616 & 0.3581 & 0.2715 \\
\multicolumn{1}{l|}{TLSH}                & 0.9526 & 0.3768 & 0.4883 & 0.6625 & 0.6201 
& 0.9304 & 0.0168 & 0.0809 & 0.2758 & 0.3260 \\
\multicolumn{1}{l|}{SSDEEP}              & 0.9393 & 0.2371 & 0.1598 & 0.6412 & 0.4944 
& 0.8050 & 0.0055 & 0.0369 & 0.2436 & 0.2728 \\

\midrule

\multicolumn{1}{l|}{\revision{\text{KEENHash$_\text{w/oFH}^{16}$}}} & \revision{0.9580} & \revision{0.6613} & \revision{0.6410} & \revision{0.7430} & \revision{0.7508} 
& \revision{0.9343} & \revision{0.0261} & \revision{0.4226} & \revision{0.3972} & \revision{0.4451} \\

\multicolumn{1}{l|}{\meanpooling{} (Baseline)}            & 0.9599 & 0.7289 & 0.7369 & 0.7482 & 0.7935 
& 0.9335 & 0.0333 & 0.5263 & 0.4081 & 0.4753 \\

\multicolumn{1}{l|}{\revision{LoC}} & \revision{0.9595} & \revision{0.7438} & \revision{0.7406} & \revision{0.7577} & \revision{0.8004} 
& \revision{0.9336} & \revision{0.0339} & \revision{0.5548} & \revision{0.4123} & \revision{0.4837} \\

\multicolumn{1}{l|}{\revision{NoS}} & \revision{0.9604} & \revision{0.7610} & \revision{\textbf{0.7972}} & \revision{0.7486} & \revision{0.8168} 
& \revision{0.9330} & \revision{0.0341} & \revision{\textbf{0.6604}} & \revision{0.4022} & \revision{0.5074} \\

\midrule

\multicolumn{1}{l|}{\keenhashstructural{16}} & 0.9602 & 0.7247 & 0.7270 & 0.7519 & 0.7910 
& \textbf{0.9363} & 0.0322 & 0.5467 & \textbf{0.4144} & 0.4824 \\
\multicolumn{1}{l|}{\keenhashsemantic}       & \textbf{0.9608} & \textbf{0.7628} & 0.7911 & \textbf{0.7581} & \textbf{0.8182} 
& 0.9361 & \textbf{0.0346} & 0.6531 & 0.4126 & \textbf{0.5091} \\

\bottomrule
\end{tabular}}
\end{minipage} 
\end{table}

\noindent $\bullet$ \textbf{String-based Method.} We introduce 1 string-based method to demonstrate its vulnerability to simple string-based attacks (the reason for excluding it from the following RQs). 
The string-based method (\minhashstring) extracts strings from a binary through command \texttt{strings} and uses MinHash~\cite{broder1997resemblance} to generate 128 hash values as its representation, in the Jaccard similarity space. Real-world malware usually avoids meaningful string literals, making challenges to the string-based analysis systems. Mirai~\cite{antonakakis2017understanding} is a famous botnet family, targeting various kinds of IoT devices for DDoS attacks. The initial version has been open-sourced since 2016~\cite{mirai}. By analyzing the source code of Mirai, we discover that Mirai encrypts almost all of its strings. Therefore, we substitute the secret key originally used by Mirai and encrypt all strings in the source code using the same encryption strategy, where the program behavior remains unchanged. Furthermore, we compile both original (Mirai$_N$) and newly obfuscated (Mirai$_O$) versions across 9 architectures, 5 optimization levels, and 2 Mirai options, obtaining 90 binaries per version. These binaries are taken as the query dataset to search against the IoT repository which contains Mirai variants.

According to the results shown in Table \ref{tab:programclonesearchmirai}, apart from \minhashstring{}, all other methods show consistent performance in mAP@100 and mP@100 against both Mirai$_N$ and Mirai$_O$ since they do not mainly rely on string features. However, turn to \minhashstring{}, its mAP@100 and mP@100 decrease by 80.91\% and 89.6\% from Mirai$_N$ to Mirai$_O$. The reason for \minhashstring{}, against Mirai$_O$, still achieving a 0.1040 mP@100 is due to the unobfuscated strings added during the compilation process. Such a result alerts that \minhashstring{} is not a reliable system since adversaries can easily breach it using simple string encryption strategies for hiding the original information of string literals. Therefore, we exclude string-based methods from the comparison in RQ2 to RQ5.

\noindent \textbf{Result.} Table \ref{tab:programclonesearch} presents the program clone search results on the respective datasets, including their average results. \keenhash{} (i.e., \keenhashstructural{16} and \keenhashsemantic{}) methods outperform all other structure-based methods, by averages of at least 12.69\% and 8.66\% in mAP@100 and mP@100 (i.e., \psso{}), respectively. Furthermore, \keenhash{} demonstrates more distinct advantages on the two benign datasets with massive code reuse (see below). For instance, \psso{} on BinaryCorp scores a rather low mAP@100 at only 0.3212 (e.g., lower than \keenhashstructural{16}'s 0.7247). Even within malware datasets, \keenhash{} continues to outperform these methods in both metrics. Since these structure-based methods, depending only on simple features (e.g., \psso{}'s), extract no semantic information, the effectiveness of similarity evaluation between binaries can be significantly impacted across compilation environments (e.g., O0 vs. O3~\cite{ding2019asm2vec}) and (malware) variants. Instead, \keenhash{} leverages semantics in pseudo functions, mitigating the impact of these factors.

\noindent $\bullet$ \textbf{\revision{Ablation Study.}} \revision{Comparing \keenhashstructural{16} with \text{KEENHash$_\text{w/oFH}^{16}$}, Table \ref{tab:programclonesearch} shows that the former outperforms the latter across all datasets. For instance, \keenhashstructural{16} achieves 0.7270 mAP@100 on BinKit$_{N}$, significantly surpassing \text{KEENHash$_\text{w/oFH}^{16}$}'s by 8.6\%. As indicated by Sec. \ref{sec:eval:functionmatching}, smaller cluster sizes $n$ of K-Means affect the function matching results, which in turn impacts the quality of the generated program embeddings. These two experimental results demonstrate the necessity of a well-configured K-Means (i.e., $n$=$2^{22}$) in conjunction with Feature Hashing.}

\revision{Regarding LoC and NoS, in general, both LoC and NoS perform better than Mean Pooling (without any feature), indicating that either of them has a positive contribution. For example, LoC and NoS get 0.7438 and 0.7610 mAP@100, higher than Mean Pooling's 0.7289.
In addition, \keenhashsemantic{} with both features is generally better than LoC and NoS. For instance, on MLWMC, \keenhashsemantic{} achieves 0.7581 mAP@100, outperforming both LoC and NoS.
Therefore, we can conclude that the introduction of both LoC and NoS is useful for enhancing \keenhashsemantic{} (which obtains the best results on average, compared with either LoC or NoS).
}

\revision{As for Mean Pooling, we analyze and discuss it later as it relates to RQ2 and RQ3.}

\noindent $\bullet$ \textbf{Massive Code Reuse.} The two \keenhash{} and the pooling methods exhibit similar performance on IoT and MLWMC while \keenhashsemantic{} performs significantly better than others on the benign datasets (e.g., 5.42\% and 12.68\% improvements in mAP@100 and mP@100 to Mean Pooling for BinKit$_{N}$). 
A common feature of these \revision{\textit{two}} datasets is that multiple non-same-class binaries can be compiled (same compile environment) from a single project, and they share (i.e., massive code reuse) a large number of the common functions (e.g., around 72\%/63\%, in Jaccard similarity, between \texttt{cp} with O3 and \texttt{ln} with O3/\texttt{cp} with O0 in \texttt{Coreutils-8.29}. \keenhashsemantic{} is better in distinguishing these kinds of cases)~\cite{gnulinuxcommunity, archlinux, archuserrepository}, leading to difficulty for BCSA methods to distinguish them under the two datasets (\textit{across compile environments}).
Thus, only maximizing the unique features can more effectively distinguish them (Sec. \ref{sec:semanticsbasedembeddinggeneration}). Therefore, such a result illustrates the advantages of integrating function semantics with their intrinsic information to distinguish binaries with massive code reuse by maximizing such feature function semantics. Furthermore, though \keenhashstructural{16} has a similar performance to Mean Pooling on the 4 datasets, it shows significantly greater robustness across compile environments and under code obfuscation (Table \ref{tab:programclonesearchfinegranularity} and \ref{tab:programclonesearchobfuscation}).

\begin{table}[t]
\begin{minipage}[t]{0.55\textwidth}
\centering
\vspace{0.2em}
\caption{The performance of \keenhash{} on program clone search across compile environments on BinKit$_N$ query and repository. The second/third row represents query/repository (samples) with specific compile environments.}

\scalebox{0.45}{\begin{tabular}{cc|c|c|c|c|c|c|c|c}
\toprule
\multirow{3}{*}{\textbf{Method}} & \multirow{3}{*}{\textbf{Metric}} & \multicolumn{4}{c|}{\textbf{Optimization}} & \textbf{Compiler} & \multicolumn{3}{c}{\textbf{Architecture}}                                                           \\

\cmidrule{3-10}

                        &                         & O0 & O1 & O2 & O3                & GCC      & ARM                                      & MIPS & x86  \\
                        &                         & O1 & O2 & O3 & O0                & Clang    & MIPS & x86                                      & ARM  \\
\midrule

\multicolumn{1}{l}{\multirow{2}{*}{\psso}}      & \multicolumn{1}{l|}{mAP@$k$} & 0.4833 & 0.3606 & 0.4486 & 0.1475 & 0.3344 & 0.3179 & 0.2152 & 0.4491 \\
\cmidrule{2-10}
                        & \multicolumn{1}{l|}{mP@$k$}  & 0.1712 & 0.1282 & 0.1334 & 0.0727 & 0.1572 & 0.0916 & 0.1034 & 0.1644 \\

\midrule

\multicolumn{1}{l}{\multirow{2}{*}{Vhash}}      & \multicolumn{1}{l|}{mAP@$k$} & 0.2969 & 0.2778 & 0.3083 & 0.2655 & 0.0095 & 0.0152 & 0.0078 & 0.0086 \\
\cmidrule{2-10}
                        & \multicolumn{1}{l|}{mP@$k$}  & 0.0232 & 0.0229 & 0.0245 & 0.0222 & 0.0045 & 0.0048 & 0.0034 & 0.0040 \\

\midrule

\multicolumn{1}{l}{\multirow{2}{*}{TLSH}}      & \multicolumn{1}{l|}{mAP@$k$} & 0.0682 & 0.2955 & 0.4046 & 0.0735 & 0.1210 & 0.0547 & 0.0515 & 0.0814 \\
\cmidrule{2-10}
                        & \multicolumn{1}{l|}{mP@$k$}  & 0.0226 & 0.0479 & 0.0489 & 0.0268 & 0.0446 & 0.0237 & 0.0255 & 0.0375 \\

\midrule

\multicolumn{1}{l}{\multirow{2}{*}{SSDEEP}}      & \multicolumn{1}{l|}{mAP@$k$} & 0.0540 & 0.0864 & 0.1050 & 0.0481 & 0.0265 & 0.0584 & 0.0320 & 0.0510 \\
\cmidrule{2-10}
                        & \multicolumn{1}{l|}{mP@$k$}  & 0.0144 & 0.0184 & 0.0172 & 0.0165 & 0.0163 & 0.0187 & 0.0139 & 0.0181 \\

\midrule

\multicolumn{1}{l}{\multirow{2}{*}{\meanpooling{} (Baseline)}}      & \multicolumn{1}{l|}{mAP@$k$} & 0.7387 & 0.6956 & 0.7077 & 0.7155 & 0.7558 & 0.7095 & 0.7263 & 0.6824 \\
\cmidrule{2-10}
                        & \multicolumn{1}{l|}{mP@$k$}  & 0.3859 & 0.3701 & 0.3627 & 0.3791 & 0.5171 & 0.3865 & 0.3648 & 0.3421 \\

\midrule

\multicolumn{1}{l}{\multirow{2}{*}{\keenhashstructural{16}}}      & \multicolumn{1}{l|}{mAP@$k$} & 0.7841 & 0.7569 & 0.7293 & 0.8072 & 0.8125 & 0.7870 & 0.7843 & 0.7487 \\
\cmidrule{2-10}
                        & \multicolumn{1}{l|}{mP@$k$}  & 0.4571 & 0.4350 & 0.4051 & 0.4654 & 0.6106 & 0.4385 & 0.4144 & \textbf{0.4122} \\

\midrule

\multicolumn{1}{l}{\multirow{2}{*}{\keenhashsemantic}}      & \multicolumn{1}{l|}{mAP@$k$} & \textbf{0.8195} & \textbf{0.7817} & \textbf{0.7684} & \textbf{0.8080} & \textbf{0.8245} & \textbf{0.7958} & \textbf{0.8024} & \textbf{0.7640} \\
\cmidrule{2-10}
                        & \multicolumn{1}{l|}{mP@$k$}  & \textbf{0.4688} & \textbf{0.4613} & \textbf{0.4364} & \textbf{0.4689} & \textbf{0.6330} & \textbf{0.4620} & \textbf{0.4373} & 0.4100 \\

\bottomrule
\end{tabular}}

\label{tab:programclonesearchfinegranularity}
\end{minipage}%
\quad
\begin{minipage}[t]{0.41\textwidth}
\vspace{0em}
\centering
\caption{The performance of \keenhash{} on program clone search against code obfuscation on BinKit$_N$ repository and BinKit$_O$ query (samples) with specific compile environments.}

\scalebox{0.45}{\begin{tabular}{cc|c|c|c|c|c}
\toprule
\multirow{2}{*}{\textbf{Method}} & \multirow{2}{*}{\textbf{Metric}} & SUB & BCF & FLA & ALL & $O$ \\

    &   & $N$ & $N$ & $N$ & $N$ & $N$   \\
\midrule

\multicolumn{1}{l}{\multirow{2}{*}{\psso}}      & \multicolumn{1}{l|}{mAP@$k$} & 0.6675 & 0.3727 & 0.2085 & 0.1231 & 0.3483 \\
\cmidrule{2-7}
                        & \multicolumn{1}{l|}{mP@$k$}  & 0.3043 & 0.2170 & 0.1315 & 0.0867 & 0.1872 \\

\midrule

\multicolumn{1}{l}{\multirow{2}{*}{Vhash}}      & \multicolumn{1}{l|}{mAP@$k$} & 0.0034 & 0.0068 & 0.0048 & 0.0070 & 0.2176 \\
\cmidrule{2-7}
                        & \multicolumn{1}{l|}{mP@$k$}  & 0.0023 & 0.0041 & 0.0033 & 0.0036 & 0.0536 \\

\midrule

\multicolumn{1}{l}{\multirow{2}{*}{TLSH}}      & \multicolumn{1}{l|}{mAP@$k$} & 0.3776 & 0.0899 & 0.0580 & 0.0228 & 0.1395 \\
\cmidrule{2-7}
                        & \multicolumn{1}{l|}{mP@$k$}  & 0.0778 & 0.0361 & 0.0232 & 0.0096 & 0.0373 \\

\midrule

\multicolumn{1}{l}{\multirow{2}{*}{SSDEEP}}     & \multicolumn{1}{l|}{mAP@$k$} & 0.1062 & 0.0627 & 0.0338 & 0.0145 & 0.0552 \\
\cmidrule{2-7}
                        & \multicolumn{1}{l|}{mP@$k$} & 0.0393 & 0.0325 & 0.0188 & 0.0119 & 0.0260 \\

\midrule

\multicolumn{1}{l}{\multirow{2}{*}{Mean Pooling (Baseline)}}      & \multicolumn{1}{l|}{mAP@$k$} & 0.7673 & 0.6359 & 0.4377 & 0.2608 & 0.5312 \\ 
\cmidrule{2-7}
                        & \multicolumn{1}{l|}{mP@$k$}  & 0.5227 & 0.5178 & 0.3825 & 0.2281 & 0.4166 \\ 

\midrule

\multicolumn{1}{l}{\multirow{2}{*}{\keenhashstructural{16}}}      & \multicolumn{1}{l|}{mAP@$k$} & 0.7757 & \textbf{0.8207} & \textbf{0.7735} & \textbf{0.7070} & \textbf{0.7704} \\
\cmidrule{2-7}
                        & \multicolumn{1}{l|}{mP@$k$}  & 0.5679 & \textbf{0.6789} & \textbf{0.5915} & \textbf{0.5593} & \textbf{0.6006} \\

\midrule

\multicolumn{1}{l}{\multirow{2}{*}{\keenhashsemantic}}      & \multicolumn{1}{l|}{mAP@$k$} & \textbf{0.8026} & 0.7353 & 0.5032 & 0.2726 & 0.5849 \\
\cmidrule{2-7}
                        & \multicolumn{1}{l|}{mP@$k$}  & \textbf{0.6428} & 0.6350 & 0.4427 & 0.2483 & 0.4974 \\

\bottomrule
\end{tabular}}
\label{tab:programclonesearchobfuscation}
\end{minipage} 
\end{table}

\noindent $\bullet$ \textbf{Robustness across Compile Environments.} We evaluate the robustness of \keenhash{} across compile environments (see Table \ref{tab:programclonesearchfinegranularity}) on BinKit$_{N}$. Here we denote <$x$, $y$> (e.g., <O3, O0>) as that the new query and repository contain only the binaries with options $x$ and $y$ from the original ones, respectively. Thus, the primary differences between the new query and the repository are focused on $x$ and $y$. Here, we only list the representative results in Table \ref{tab:programclonesearchfinegranularity}, where the second/third-row option represents query/repository. The results show that \keenhash{} methods outperform all other structure-based methods across all options, especially on <O3, O0>. As mentioned earlier, simple features can be significantly affected by compilation environments, weakening the effectiveness of them. Furthermore, both \keenhash{} methods significantly outperform the pooling method (e.g., 9\% and 8\% improvements in mAP@100 and mP@100 on <O3, O0>), illustrating that direct integrating on function semantics exhibits poorer robustness.



\mybox{\textbf{Answer 2:} Both \keenhash{} methods significantly outperform state-of-the-art methods across 4 datasets in terms of performance by an average of at least 12.69\% in mAP@100 on program clone search. In addition, \keenhashsemantic{} is more effective than \keenhashstructural{16} and Mean Pooling in the massive code reuse scenario. Both \keenhash{} methods also show greater robustness across compilation environments than others including Mean Pooling.}

\subsection{RQ3: \keenhash{} against Code Obfuscation}\label{sec:eval:againstobfuscation}

\noindent \textbf{Motivation.} Code obfuscation is the process of modifying binaries to make them no longer useful to hackers while maintaining them fully functional. On the contrary, it can interfere with existing BCSA methods. In this RQ, we assess the robustness of \keenhash{} against code obfuscation.  

\noindent \textbf{Approach.} We use BinKit$_{O}$'s query to retrieve binaries from BinKit$_{N}$'s repository in an obfuscation vs. normal scenario. In BinKit$_{O}$, SUB transforms fragments of assembly code to their equivalent form through predefined rules; BCF modifies the CFGs of functions by adding extensive irrelevant basic blocks; FLA changes the original CFG using a complex hierarchy of new conditions as switches; and, ALL combines all obfuscations above. The evaluation metrics include mAP@100 and mP@100.

\noindent \textbf{Result.} Table \ref{tab:programclonesearchobfuscation} presents the retrieving results against code obfuscation where 'O'/‘N’ represents the complete query or repository of BinKit$_{O}$/BinKit$_{N}$. According to the results, \keenhash{} methods consistently outperform other structure-based methods in all scenarios (e.g., \keenhashstructural{16} outperforms \psso{} by 58\% and 47\% in mAP@100 and mP@100 on <ALL, $N$>). In obfuscation scenarios, semantics in pseudo functions still distinguish binaries better than simple features.

Furthermore, in general, \keenhashstructural{16} has better robustness with 0.7704 mAP@100 and 0.6006 mP@100 on <$O$, $N$> than \keenhashsemantic{}'s 0.5849 and 0.4974. Although code obfuscation can impact the effectiveness of function embeddings, especially for <ALL, $N$>, our K-Means model mitigates this by transforming function similarity into a matching problem (i.e., 0 or 1) and still capturing enough relationships between normal and obfuscated functions. Thereby, it can reduce the impact of similarities among normal and obfuscated functions. On <SUB, $N$>, \keenhashsemantic{} is slightly better since the impact of SUB on function embeddings is relatively weak. Additionally, the two \keenhash{} methods perform significantly better than Mean Pooling in both metrics (e.g., 23.92\% and 18.4\% improvements in mAP@100 and mP@100 on <$O$, $N$> for \keenhashstructural{16}), demonstrating again the advantages of structure and semantics-based program embedding generation of \keenhash{}.




\mybox{\textbf{Answer 3:} \keenhash{} is more robust against code obfuscation than state-of-the-art methods by at least 23.66\% - 42.21\% in mAP@100 on <$O$, $N$>. In addition, the performance of \keenhashstructural{16} is generally better than \keenhashsemantic{}. Both \keenhash{} methods are also more robust than Mean Pooling by 5.37\% - 23.92\% in mAP@100 on <$O$, $N$>.}

\subsection{RQ4: \keenhash{} on Larger-scale Repository}\label{sec:eval:all}

\begin{table*}
\caption{Program clone search against all datasets of IoT, BinaryCorp, BinKit$_N$, MLWMC, BinKit$_O$, Mirai$_N$, and Mirai$_O$. The query datasets are shown in the table and the repository is the merge of repositories of IoT, BinKit$_N$, BinaryCorp, and MLWMC.}
\centering
\scalebox{0.50}{\begin{tabular}{cccccccc|ccccccc}
\toprule

\multirow{2}{*}{\textbf{Method}}   & \multicolumn{7}{c|}{\textbf{mAP@100}} & \multicolumn{7}{c}{\textbf{mP@100}}     \\

\cmidrule{2-15}

                    & \multicolumn{1}{c}{{IoT}} & \multicolumn{1}{c}{{BinaryCorp}} & \multicolumn{1}{c}{{BinKit$_N$}} & \multicolumn{1}{c}{{MLWMC}} & \multicolumn{1}{c}{{BinKit$_O$}} & \multicolumn{1}{c}{{Mirai$_N$}} & \multicolumn{1}{c|}{{Mirai$_O$}} & \multicolumn{1}{c}{{IoT}} & \multicolumn{1}{c}{{BinaryCorp}} & \multicolumn{1}{c}{{BinKit$_N$}} & \multicolumn{1}{c}{{MLWMC}} & \multicolumn{1}{c}{{BinKit$_O$}} & \multicolumn{1}{c}{{Mirai$_N$}} & \multicolumn{1}{c}{{Mirai$_O$}} \\
\midrule

\multicolumn{1}{l|}{\psso}  & 0.9144 & 0.3136 & 0.6798 & 0.7153 & 0.3256 & 0.8781 & 0.8430 & 
0.8436 & 0.0123 & 0.2779 & 0.3623 & 0.1615 & 0.8470 & 0.8183 \\

\midrule

\multicolumn{1}{l|}{Vhash}               & 0.6812 & 0.5651 & 0.3414 & 0.7375 & 0.2179 & 0.3931 & 0.3931 & 
0.6156 & 0.0206 & 0.0604 & 0.3585 & 0.0524 & 0.3917 & 0.3917 \\
\multicolumn{1}{l|}{TLSH}                & 0.9438 & 0.3769 & 0.4888 & 0.6801 & 0.1362 & 0.6692 & 0.6732 & 
0.9148 & 0.0168 & 0.0797 & 0.3206 & 0.0346 & 0.6555 & 0.6661 \\
\multicolumn{1}{l|}{SSDEEP}              & 0.7045 & 0.2375 & 0.1606 & 0.6413 & 0.0516 & 0.0778 & 0.0 & 
0.2258 & 0.0054 & 0.0375 & 0.2430 & 0.0222 & 0.0066 & 0.0 \\

\midrule

\multicolumn{1}{l|}{\keenhashstructural{16}} & 0.9599 & 0.7243 & 0.7270 & 0.7515 & \textbf{0.7704} & \textbf{0.9997} & \textbf{0.9999} & 
\textbf{0.9354} & 0.0322 & 0.5466 & \textbf{0.4141} & \textbf{0.6004} & \textbf{0.9996} & \textbf{0.9997} \\
\multicolumn{1}{l|}{\keenhashsemantic}       & \textbf{0.9609} & \textbf{0.7627} & \textbf{0.7911} & \textbf{0.7553} & 0.5666 & 0.9937 & 0.9939 & 
0.9352 & \textbf{0.0346} & \textbf{0.6530} & 0.4116 & 0.4910 & 0.9883 & 0.9873 \\

\bottomrule
\end{tabular}}
\label{tab:programclonesearchallbinaries}
\end{table*}

\noindent \textbf{Motivation.} The comparisons should not be affected greatly by the scale and distributions of binaries. In this RQ, we evaluate \revision{the robustness of} \keenhash{} in distinguishing binaries on a larger-scale dataset\revision{, simulating the diversity of data distributions in real-world scenarios (e.g., across Linux/Windows and benign/malicious binaries)}.

\noindent \textbf{Approach.} We merge the repositories of IoT, BinKit$_N$, BinaryCorp, and MLWMC in sequence to form a larger repository, where the order of samples within each repository is maintained. Furthermore, we use 7 query datasets from all previous experiments. In total, there are 171,075 and 31,230 binaries in the repository and query, respectively, indicating 5.3 billion similarity evaluations (\textit{a typical evaluation workload within one day} based on VirusTotal-related reports~\cite{virustotal, van2023deep, dambra2023decoding}). The metrics include mAP@100 and mP@100.

\noindent \textbf{Result.} The experimental results are shown in Table \ref{tab:programclonesearchallbinaries}. By Cliff’s Delta effect size~\cite{macbeth2011cliff} between the pair of results (i.e., two metrics) of respective repositories and the larger repository to each method, the measured effect sizes show that only \psso{} and SSDEEP have small and large differences, while others have negligible differences, in general. Where SSDEEP shows a significant drop in both metrics on the IoT query (\psso{} has a drop in mP@100 by 5.6\%). Furthermore, by investigating Mirai$_N$ and Mirai$_O$, the 4 structure-based methods, with simple features, have an obvious performance decrease compared to the results in Table \ref{tab:programclonesearchmirai}. In contrast, \keenhash{} methods maintain consistent performance. As a result of the dataset merge, the capability of structure-based methods to differentiate Mirai from binaries in BinaryCorp, BinKit$_N$, and MLWMC is negatively affected, demonstrating that they can be potentially and significantly affected by the scale and distributions of binaries. Overall, across the 7 queries, two \keenhash{} methods can maintain consistent performance (only nearly consistent for \keenhashsemantic{} to BinKit$_O$ due to code obfuscation to the effectiveness of function embeddings), while the 4 structure-based methods always have obvious and slight (e.g., IoT) degradation to some of these queries.

We further plot the performance of program clone search across multiple $k \in [1, 100]$ in Fig. \ref{fig:programclonesearchall}, where the query is the merged one of all the 7 queries. The results show that in general, \keenhash{} can achieve at least 0.7647 mAP@100 and 0.3858 mP@100 and outperforms other structure-based methods (by at least 23.16\% and 13.79\% in $k=100$, i.e., \psso{}) across different $k$, indicating its enhanced capacity on large-scale BCSA. Moreover, we highlight that \keenhashstructural{16} and \keenhashsemantic{} costs only 395.83 and 90.31 seconds with CPU (48 cores), respectively, for 5.3 billion similarity evaluations (i.e., scalable in large-scale scenarios).

\mybox{\textbf{Answer 4:} In general, \keenhash{} outperforms other structure-based methods by at least 23.16\% in mAP@100, and shows greater robustness, against a larger-scale dataset. Additionally, it is scalable for large-scale BCSA scenarios. For instance, in the scenario with 5.3 billion similarity evaluations, \keenhash{} takes at most 395.83 seconds.}

\subsection{RQ5: \keenhash{} on Malware Detection}\label{sec:eval:malware}

\noindent \textbf{Motivation.} In previous RQs, we have demonstrated the superiority of \keenhash{} from aspects including effectiveness and robustness. In this RQ, we show how \keenhash{} can be helpful in a more specific and critical large-scale BCSA security application: malware detection.

\noindent \textbf{Approach.} We leverage the merged repository used in RQ4 as the malware and benign binary database. Additionally, we perform malware detection (i.e., benign or malicious) by leveraging IoT, Mirai$_{N\text{/}O}$, and BinKit$_N$ queries against the merged repository from the malware security aspect. We use $K$-NN as the binary classifier and the numbers of false negatives (i.e., misclassified as benign samples) and false positives (i.e., misclassified as malicious samples) as the metrics.

\begin{figure}[t]
\begin{minipage}[t]{0.48\textwidth}
\vspace{0em}
\centering
    \includegraphics[width=1.0\textwidth]{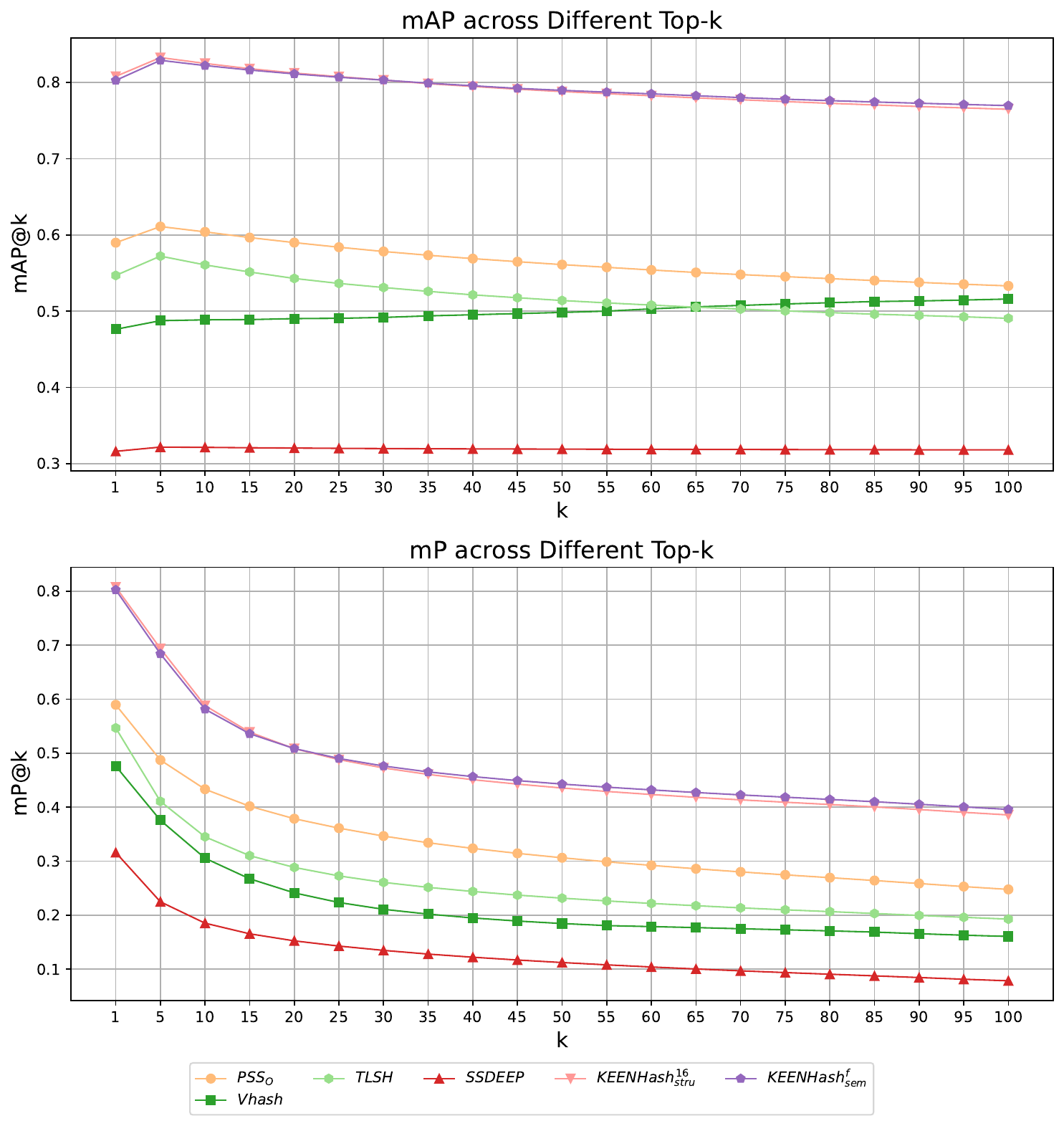}
    \caption{Program clone search on all binaries.}  
    \label{fig:programclonesearchall}
\end{minipage}%
\quad
\begin{minipage}[t]{0.48\textwidth}
\vspace{3em}
\captionof{table}{Malware detection against Mirai$_{N\text{/}O}$ (malicious), IoT (malicious), and BinKit$_N$ (benign) dataset. Numbers represent how many binaries are \emph{misclassified} by the BCSA methods in different $K$-NN settings.}
\centering
\scalebox{0.53}{\begin{tabular}{lccc|ccc|ccc}
\toprule
\multicolumn{1}{c|}{\multirow{3}{*}{\textbf{Method}}} & \multicolumn{3}{c|}{\textbf{Mirai$_{N\text{/}O}$}} & \multicolumn{3}{c|}{\textbf{IoT}} &  \multicolumn{3}{c}{\textbf{BinKit$_N$}} \\
\cmidrule{2-10}

& \multicolumn{6}{|c|}{\textbf{False Negative Number}} & \multicolumn{3}{c}{\textbf{False Positive Number}} \\

\cmidrule{2-10}
\multicolumn{1}{c|}{}                        & $K$=1 & $K$=3 & $K$=5           & $K$=1 & $K$=3  & $K$=5  & $K$=1 & $K$=3  & $K$=5 \\
\midrule
\multicolumn{1}{l|}{\psso \textcolor{red}{\ding{55}}}                  & 30  & 18  & 12            & 29  & 45   & 51   &  5 & 16   &  31      \\
\midrule
\multicolumn{1}{l|}{Vhash \textcolor{red}{\ding{55}}}                  & 40  & 40  & 60            & 34  & 46   & 59     & 2  &  4  &  5   \\
\multicolumn{1}{l|}{TLSH \textcolor{red}{\ding{55}}}                   & 50  & 52  & 52            & 14  & 18   & 17     &  1 &  1  &   1    \\
\multicolumn{1}{l|}{SSDEEP \textcolor{red}{\ding{55}}}                 & \textbf{0}   & 174 & 174           & \textbf{0}   & 1143 & 1514   & 3856  & 1   &   1    \\

\midrule
\multicolumn{1}{l|}{\keenhashstructural{16} \textcolor{green}{\ding{51}}}& \underline{\textbf{0}}   & \underline{\textbf{0}}   & \underline{\textbf{0}}  & \underline{\textbf{0}} & \underline{\textbf{0}}  & \underline{\textbf{0}}  & \underline{\textbf{0}} &  \underline{\textbf{0}} & \underline{\textbf{0}}   \\
\multicolumn{1}{l|}{\keenhashsemantic \textcolor{green}{\ding{51}}}      & \underline{\textbf{0}} & \underline{\textbf{0}} & \underline{\textbf{0}}  & \underline{\textbf{0}} & \underline{\textbf{0}}  & \underline{\textbf{0}}  & \underline{\textbf{0}} & \underline{\textbf{0}} & \underline{\textbf{0}}  \\
\bottomrule
\end{tabular}}
\label{tab:malwaredetection}

\end{minipage} 
\end{figure}

\noindent \textbf{Result.} As shown in Table \ref{tab:malwaredetection}, \keenhash{} (both structural and semantic) outputs \emph{zero} misclassification in all three datasets across malicious and benign, in all $K$-NN settings. In comparison, all other BCSA methods can make mistakes, even for the widely-recognized security method Vhash from VirusTotal.
%
%
%
We identified two primary causes of misclassified cases: (1) the query binary belongs to a less common architecture (e.g., m68k), resulting in significant code differences compared to the repository ones; and (2) the query binary is a variant of repository ones, but with substantial changes in code functions (e.g., much more functions and invocations), which simple features (e.g., spectrum of call graph) fail to capture them.
Thanks to the capability to capture function semantics and represent binaries, \keenhash{} can accurately classify these challenging malware samples, while other methods with simple features struggle with such variations, resulting in false negatives.
%
%
For false positives, the shortcomings of these methods across compilation environments (see Sec. \ref{sec:eval:programclonesearch}) lead to prioritizing the retrieval of malicious binaries in certain cases.

\mybox{\textbf{Answer 5:} Thanks to the powerful LLM-based function embedding and effective program representation, \keenhash{} demonstrates superior performance in the large-scale BCSA scenario of malware detection, and significantly outperforms previous methods including the widely-recognized method Vhash from VirusTotal.}

%% file: secs/discussion.tex
\section{Limitations}\label{sec:limitations}

\keenhash{} suffers from the limitations to the property of static analysis including decompilation. For example, packed binaries and binaries with payloads are difficult to decompile accurately to obtain precise function results, thus affecting its performance. Furthermore, substantial variations in or among functions warrant caution, like significant source code revisions (e.g., code refactoring), aggressive inter-procedural compiler optimizations (e.g., function inlining and link-time optimizations), and great inter or inner-procedural code obfuscations (e.g., function merging). These variations may disrupt the similarity between program structures or affect the performance of our function embedding model. In the future, we plan to systematically evaluate the impact of such variations and propose corresponding mitigation strategies. In addition, the maximum length of input tokens to our function embedding model is limited due to space consideration~\cite{vaswani2017attention}. This limitation can be overcome with new LLM architectures introduced in the future.



%% file: secs/relatedwork.tex
\section{Related Work}\label{sec:relatedwork}

\noindent \textbf{Function-level Similarity Analysis.} Recently, there has been a tremendous increase in the popularity of (binary) function-level similarity analysis~\cite{wang2022jtrans, kim2022revisiting, ding2019asm2vec, wang2024clap, wang2023sem2vec, haq2021survey, yu2020order, yu2020codecmr, ming2017binsim}. Considering performance and scalability, these function representation techniques mainly focus on static analysis and deep learning. Gemini~\cite{xu2017neural} extracts crafted features of each basic block from Genius~\cite{feng2016scalable} and employs GNN to learn the representations of CFG to functions.
%
%
%
jTrans~\cite{wang2022jtrans} leverages a transformer-based method with a jump-aware representation of the analyzed binary functions and a newly-designed pre-training task to generate embeddings, encoded with CFG information. VulHawk~\cite{luo2023vulhawk} proposes an intermediate representation function model with the language model and GCN, followed by an entropy-based adapter to transfer function embedding space from different file environments into the same one to alleviate the differences caused by various file environments. CLAP~\cite{wang2024clap} boosts superior transfer learning capabilities by effectively aligning binary code with their semantics explanations. However, it is difficult to directly adapt them in program-level BCSA due to the scalability problem.




\noindent \textbf{Program-level Similarity Analysis.} There are few recent studies related to program-level BCSA~\cite{hu2013mutantx, benoit2023scalable, duan2020deepbindiff, haq2021survey}. SSDEEP~\cite{kornblum2006identifying} is a fuzzy hashing technique based on Context Triggered Piecewise Hashing (CTPH) to hash files into hash strings. TLSH~\cite{oliver2013tlsh} is a fuzzy hashing method based on k-skip N-gram features, followed by LSH to hash feature counts into a vector with 128 buckets. Vhash~\cite{virustotal} is an in-house similarity clustering algorithm, based on a simple structural feature hash. \psso{}~\cite{benoit2023scalable} is a spectral-based method that represents programs through their spectrums of call graphs and the edge counts of CFGs. These methods are limited to their performance on large-scale BCSA. There are also some studies evaluating similarity based on dynamic analysis and symbolic execution~\cite{anderson2011graph, jang2011bitshred, ming2017binsim, luo2017semantics}. However, such methods are limited to the explorable execution space and the runtime overhead of the representation generation in large-scale scenarios. Furthermore, some work~\cite{duan2020deepbindiff, diaphora, binaryai, gao2024sigmadiff} focuses on binary diffing. While these methods also fail to scale to large-scale scenarios due to the scalability problem, which is shown in our experiment.




%% file: secs/conclusion.tex
\section{Conclusion}\label{sec:conclusion}

In this paper, we propose a novel large-scale program-level BCSA hashing approach \keenhash{}, for evaluating similarities among binaries. \keenhash{} captures binaries from the dual perspectives of function matching based on K-Means and Feature Hashing, and program semantics by integrating function embeddings to generate respective compact and fixed-length program embeddings. Our experimental results demonstrate that \keenhash{} is \revision{at least} 215 times faster than the state-of-the-art function matching \revision{tools} while maintaining effectiveness. Furthermore, in a large-scale scenario with 5.3 billion similarity evaluations, \keenhash{} takes only 395.83 seconds while the previous \revision{tools} will cost \revision{at least} 56 days. We also evaluate the two \keenhash{} methods on the program clone search of large-scale BCSA across extensive datasets in a total of 202,305 binaries. Compared with 4 state-of-the-art methods, \keenhash{} outperforms all of them by at least 23.16\% and displays remarkable superiority over them in the BCSA security scenario of malware detection. Such results demonstrate the outstanding effectiveness of \keenhash{} on large-scale program-level BCSA.

%% file: secs/appendix.tex
\section{Hashed Length of Feature Hashing}\label{app:hashvaluesize}

In this section, we introduce \keenhash{}-stru with different parameter settings of hashed length for comparison. Here, we denote \keenhash{} with different settings as \keenhashstructural{y} where $y \in \{10, 12, 14, 16, 18\}$ represents the Feature Hashing size (length) of $2^y$. We evaluate the performance of \keenhashstructural{y} through the task in Sec. \ref{sec:eval:programclonesearch} (the results of other tasks show a similar trend here). The experimental results are shown in Table. \ref{tab:programclonesearchhashsize}.  On average, the performance of \keenhashstructural{y}{} improves in tandem with increases in the hashed length $2^y$. In addition, Cliff’s Delta effect size~\cite{macbeth2011cliff} measures that the effect size between any pair of results of \keenhashstructural{y}, across the 4 datasets, remains below 0.1, indicating a negligible difference. Although the performance increases on these datasets are minimal, we consider that the hashed length $2^y$ should be sufficiently large within an appropriate range to minimize the impact of the number of functions in binaries on \keenhash{}-stru for even larger-scale scenarios (e.g., 10 million binaries). Therefore, as mentioned in Sec. \ref{sec:structurebasedembeddinggeneration}, we set the hashed length to $2^{16}$ (i.e., $y=16$). 

\begin{table*}
\caption{The performance of program clone search with different hashed lengths for \keenhash{}-stru.}
\centering
\scalebox{0.7}{\begin{tabular}{cccccc|ccccc}
\toprule

\multirow{2}{*}{\textbf{Method}}   & \multicolumn{5}{c|}{\textbf{mAP@100}} & \multicolumn{5}{c}{\textbf{mP@100}}     \\

\cmidrule{2-11}

                    & \multicolumn{1}{c}{{IoT}} & \multicolumn{1}{c}{{BinaryCorp}} & \multicolumn{1}{c}{{BinKit$_N$}} & \multicolumn{1}{c}{{MLWMC}} & \multicolumn{1}{c|}{{Avg.}} & \multicolumn{1}{c}{{IoT}} & \multicolumn{1}{c}{{BinaryCorp}} & \multicolumn{1}{c}{{BinKit$_N$}} & \multicolumn{1}{c}{{MLWMC}} & \multicolumn{1}{c}{{Avg.}} \\
\midrule

\multicolumn{1}{l|}{\keenhashstructural{10}} & 0.9596 & 0.6915 & 0.7262 & 0.7424 & 0.7799 
& 0.9360 & 0.0296 & 0.5468 & 0.3961 & 0.4771 \\
\multicolumn{1}{l|}{\keenhashstructural{12}} & 0.9605 & 0.7191 & 0.7267 & 0.7502 & 0.7891 
& 0.9360 & 0.0312 & 0.5458 & 0.4105 & 0.4809 \\
\multicolumn{1}{l|}{\keenhashstructural{14}} & 0.9603 & 0.7237 & 0.7270 & 0.7518 & 0.7907 
& 0.9365 & 0.0318 & 0.5465 & 0.4135 & 0.4821 \\
\multicolumn{1}{l|}{\keenhashstructural{16}} & 0.9602 & 0.7247 & 0.7270 & 0.7519 & 0.7910 
& 0.9363 & 0.0322 & 0.5467 & 0.4144 & 0.4824 \\
\multicolumn{1}{l|}{\keenhashstructural{18}} & 0.9604 & 0.7253 & 0.7270 & 0.7518 & 0.7911 
& 0.9363 & 0.0325 & 0.5467 & 0.4150 & 0.4826 \\

\bottomrule
\end{tabular}}
\label{tab:programclonesearchhashsize}
\end{table*}

\section{Similarity Evaluation for \keenhash{}-stru}\label{app:sim}

The program embedding produced by \keenhash{}-stru is a bit-vector, with each element indicating the classification of functions for representing function matching. Generally, two popular similarity evaluation metrics are appropriate for \keenhash{}-stru: Hamming distance~\cite{douze2024faiss} and Jaccard similarity~\cite{jang2011bitshred}. In this study, we opt for the latter since the Hamming distance, quantifying absolute differences, is more likely to lead to false positives. For instance, consider programs $A$, $B$, and $C$ with 10K, 4K, and 4 numbers of 1s in their bit-vectors. $A$ and $B$ are of the same class, with the only difference being that $A$ is compiled with O0 while $B$ is compiled with O3. The hamming distance between $A$ and $B$ is at least 6K, whereas the distance between B and C is no more than $\text{4K}+\text{4}$, leading to $B$ and $C$ appearing more similar in the similarity comparison. While, Jaccard similarity incorporates proportional measures (e.g., 4K and 10K are proportionally closer than 4 and 4K), mitigating the negative effects of the absolute differences.

\section{Dataset in Detail}\label{app:datasetindetail}

This section extends the original description of the dataset used in our experiment.

\noindent \textbf{Training Dataset.} Two training datasets are included in this study for \blackding{1} the function embedding model and \blackding{2} the K-Means model, respectively. For \blackding{1}, to obtain a large number of matched source and pseudo functions, we build the automatic compilation pipeline based on ArchLinux official repositories (AOR)~\cite{archlinux} and Arch User Repository (AUR)~\cite{archuserrepository}, following the same setting in jTrans (i.e., BinaryCorp)~\cite{wang2022jtrans}. We compile all the projects through the command \texttt{makepkg}. In addition, over a period of three years, we collect open-sourced C/C++ projects from the Linux Community, along with their corresponding compiled binaries across various architectures (e.g., x86, arm, and so forth), ultimately amassing around 900K projects. Furthermore, the source functions (with matched pseudo ones) causing data leakage are excluded through sha256 (Sec. \ref{sec:1trainingphase}) from the training dataset for the evaluation of the effectiveness of our K-Means model (Sec. \ref{sec:eval:functionmatching}) and our function embedding model (Appendix \ref{app:discussion}). The source and pseudo functions and pairs are extracted through the Function Extraction (see Sec. \ref{sec:1trainingphase}). Eventually, we obtain 4.51M matched function pairs with an average of 556 tokens per function as the training dataset for the function embedding model. As for \blackding{2}, to obtain massive and diverse C/C++ source functions, we follow previous studies~\cite{tang2022towards, wu2023ossfp} to collect a large number of open-source C/C++ projects by crawling from Github~\cite{github} and GNU/Linux community~\cite{gnulinuxcommunity}. In total, 11,013 projects, including malicious ones (e.g., gh0st RAT malware~\cite{gh0st}), are obtained, containing 56M unique C/C++ source functions. A substantial source function dataset is essential for the generalization of \keenhash{}-structural.

\noindent \textbf{Test Dataset.} The test dataset is used to evaluate the performance of our K-Means model (Sec. \ref{sec:eval:functionmatching}). Specifically, we use the binary diffing dataset in DeepBinDiff~\cite{duan2020deepbindiff}. The dataset, compiled with GCC 5.4, utilizes three popular binary sets of Coreutils~\cite{coreutils}, Diffutils~\cite{diffutils}, and Findutils~\cite{findutils}, across various versions (5 versions for Coreutils, 4 versions for Diffutils, and 3 versions of Findutils) and optimization levels (O0, O1, O2, and O3). In total, there are 2,098 binaries. Moreover, the function matching ground truth is obtained through the Function Extraction in a total of 101,289 pairs of matched functions across 1,926 compared pairs of same-class binaries.

\noindent \textbf{Repository and Query Dataset.} To evaluate \keenhash{} on program clone search, we collect five datasets:

\noindent $\blacktriangleright$ \textbf{IoT.} We collect 37,657 nonpacked C/C++ (detected with DIE~\cite{die}) IoT malware samples from MalwareBazaar~\cite{malwarebazaar} across 21 malware families, spanning from January 2020 to July 2023. The malware families are obtained from VirusTotal~\cite{virustotal} reports through avclass~\cite{sebastian2016avclass, sebastian2020avclass2}, containing many notorious ones such as Mirai, Gafgyt, Tsunami, and so forth. Furthermore, each family contains at least 20 samples. We randomly divide the dataset into repository and query datasets in a 9:1 ratio and each family has at least two samples in the query.

\noindent $\blacktriangleright$ \textbf{BinaryCorp.} BinaryCorp~\cite{wang2022jtrans} dataset is crafted based on AOR and AUR. Where, AOR contains tens of thousands of diverse packages, ranging from editor, HTTP server, compiler, graphics library, cryptographic library, and so forth. AUR contains over 77,000 packages uploaded and maintained by users. Furthermore, ArchLinux provides a useful tool \texttt{makepkg} for users to build their packages from source code. Wang et al.~\cite{wang2022jtrans} choose the C/C++ project in the pipeline to build the datasets across five optimization levels of O0, O1, O2, O3, and Os. In total, 9,819 source code are collected and 45,593 distinct C/C++ binaries in x86 are generated, with 9,498 sample families. The sample family number is lower than the project number since binaries compiled (in the same compile environments) from different projects may have the same SHA256 hash values. We randomly divide the dataset into repository and query in a 7.5:2.5 ratio and each family has at least one sample in the query. Furthermore, all binaries are stripped, which is practical in real-world scenarios.

\noindent $\blacktriangleright$ \textbf{BinKit.} BinKit~\cite{kim2022revisiting} dataset is crafted from 51 GNU software packages with 235 unique C source code (i.e., sample families). The 51 GNU packages are chosen due to their popularity and accessibility as they are real-world applications that are widely used on Linux, and their source code is publicly available. The compiled binaries are also diverse along different optimization levels, compilers, architectures, and obfuscations.

\begin{itemize}
    \item \textbf{Normal (BinKit$_N$):} The normal one is compiled with 288 different compile environments for a total of 67,680 binaries to the 51 packages. It covers 8 architectures (arm, x86, mips, and mipseb, each available in 32 and 64 bits), 9 compilers (5 versions of GCC v\{4.9.4, 5.5.0, 6.4.0, 7.3.0, 8.2.0\} and 4 versions of Clang v\{4.0, 5.0, 6.0, 7.0\}), and 4 optimization levels (O0, O1, O2, and O3);

    \item \textbf{Obfuscation (BinKit$_O$):} The obfuscation one is compiled with 4 obfuscation options including instruction substitution (SUB), bogus control flow (BCF), control flow flattening (FLA), and all combined (ALL), through Obfuscator-LLVM~\cite{ieeespro2015-JunodRWM} as the compiler. Where SUB transforms fragments of assembly code to their equivalent form through predefined rules; BCF modifies the control flow graph (CFG) of functions by adding a large number of irrelevant basic blocks and branches; FLA changes the original CFG using a complex hierarchy of new conditions as switches; and, ALL combines all obfuscations above. The same architectures and 5 optimization levels (extra Os) are also covered. Therefore, a total of 37,600 binaries are generated.
\end{itemize}

The BinKit$_N$ dataset is divided randomly into repository and query in a 9:1 ratio. 10\% of the samples are randomly selected from the BinKit$_O$ dataset as the query to maintain experimental consistency in Sec. \ref{sec:eval:againstobfuscation} and \ref{sec:eval:all}. All binaries are stripped.

\noindent $\blacktriangleright$ \textbf{MLWMC.} MLWMC~\cite{dambra2023decoding} is an open PE 32 real-world malware dataset collected through VirusTotal from August 2021 to March 2022. It contains 67,000 malware samples across 670 malware families obtained from VirusTotal reports through avclass where each family contains 100 malware samples. These families belong to 13 threat categories: 36\% (282) of the families are classified as grayware, 15\% (120) as downloaders, 11\% (87) as worms, 10\% (78) as backdoors, 5\% (41) as viruses, and the remaining 23\% (62) includes ransomware, rogueware, spyware, miners, hacking tools, clickers, and dialers. In this study, we consider the 49,820 nonpacked C/C++ samples, belonging to a total of 615 malware families where each family contains at least 20 samples. Moreover, we divide the dataset in the same way as IoT.

\section{Runtime Overhead}\label{app:runtimeoverhead}

Here, we count the average time spent, across our collected 5 datasets, on each step of \keenhash{} for hashing a binary into corresponding program embeddings. In Function Extraction, the main time cost is in the decompilation process, which takes around 1 minute to decompile (1 core) one binary on average (at most 43 minutes for a binary in the size of 220MB). In Function Embedding Generation, for one binary (200 functions on average), it takes about 0.06 seconds on average (at most 12 seconds for binaries containing around 40K functions). In Program Embedding Generation, for one binary, it takes 0.3 and 0.1 seconds on average for \keenhash{}-stru (at most 35 seconds) and \keenhash{}-sem (at most 9 seconds), respectively. It is notable that these steps can be completed offline in large-scale scenarios for BCSA such as building a large-scale repository (Sec. \ref{sec:preliminary}).

\section{Function Embedding Discussion}\label{app:discussion}

As our function embedding model is the foundation of \keenhash{}, its effectiveness and ability to generalize in generating program embeddings are essential. Here, we discuss it with two state-of-the-art publically available models jTrans~\cite{wang2022jtrans} and CLAP~\cite{wang2024clap} on binary functions. We only focus on binary functions since the two models are only available on assembly code (they do not support source functions) and the performance between source and pseudo ones, for function matching, are already evaluated in Sec. \ref{sec:eval:functionmatching}. Furthermore, we employ the test dataset of BinaryCorp~\cite{wang2022jtrans} for the experiment which aligns with the one evaluated in jTrans and CLAP. The test dataset contains 2,911,846 functions across 9,351 binaries from 1,974 source code with 5 optimization levels. On average, 1 source function (i.e., class) corresponds to 5.6 binary functions.

\begin{table}
\caption{The performance of function clone search of \keenhash{}.}
\centering
\scalebox{0.7}{\begin{tabular}{c|c|c|c|c|c}
\toprule
\textbf{Method}           & \textbf{Metric} & \textbf{Top-1} & \textbf{Top-10} & \textbf{Top-50} & \textbf{Top-100}  \\
\midrule
\multirow{2}{*}{jTrans}   & \multicolumn{1}{l|}{mAP@$k$}  &0.2841 & 0.3065 & 0.2846 & 0.2714     \\
\cmidrule{2-6}
                          & \multicolumn{1}{l|}{mP@$k$}   &0.2841 & 0.1464 & 0.0640 & 0.0457 \\
\midrule

\multirow{2}{*}{CLAP}   & \multicolumn{1}{l|}{mAP@$k$}  & 0.6423 & 0.6567 & 0.6288 & 0.6186 \\
\cmidrule{2-6}
                          & \multicolumn{1}{l|}{mP@$k$}   & 0.6423 & 0.3129 & 0.1015 & 0.0648 \\
\midrule

\multirow{2}{*}{\keenhash{}} & \multicolumn{1}{l|}{mAP@$k$}  & 0.7244 & 0.7422 & 0.7202 & 0.7116 \\
\cmidrule{2-6}
                          & \multicolumn{1}{l|}{mP@$k$}   & 0.7244 & 0.3261 & 0.1162 & 0.0764 \\

\bottomrule
\end{tabular}}
\label{tab:functionclosesearch}
\end{table}

\begin{table}
\caption{Program clone search across function embedding models.}
\centering
\scalebox{0.75}{\begin{tabular}{cc|c}
\toprule

\multirow{1}{*}{\textbf{Method}}   & \multicolumn{1}{|c|}{\textbf{mAP@100}} & \multicolumn{1}{c}{\textbf{mP@100}}     \\

\cmidrule{1-3}

\multicolumn{1}{l|}{jTrans-Mean Pooling} & 0.6321 & 0.0293 \\


\multicolumn{1}{l|}{CLAP-Mean Pooling} & 0.8340 & 0.0333 \\


\multicolumn{1}{l|}{\keenhash{}-Mean Pooling} & 0.8445 & 0.0349 \\


\bottomrule
\end{tabular}}
\label{tab:programclonesearchfunctionembeddingmodel}
\end{table}

\noindent \textbf{Function Embedding Comparing.} We perform the function clone search for evaluation where the subject changes from programs to pseudo (binary) functions (Sec. \ref{sec:problemdefinition}). The test dataset is randomly divided into repository and query datasets in a 7.5:2.5 ratio where each class (if the class size $> 1$) has functions in both repository and query. The metrics include mAP@$k$ and mP@$k$. The results are shown in Table \ref{tab:functionclosesearch}. Our function embedding model (i.e., \keenhash{}) surpasses both jTrans and CLAP in terms of both mAP@$k$ and mP@$k$ metrics across $k \in \{1, 10, 50, 100\}$ on such a large and diverse dataset. The results demonstrate that our model can retrieve more same-class functions that are also ranked higher compared to jTrans and CLAP, underscoring the effective discriminative capacity of our function embedding model in generating pseudo function embeddings. For instance, our model achieves 0.7116 and 0.0764 in mAP@100 and mP@100, outperforming jTrans by 44.02\% and 3.07\%, and CLAP by 9.30\% and 1.16\%.

\noindent \textbf{Program Embedding Ablation Study.} We leverage the function embeddings produced through \keenhash{}, jTrans, and CLAP to generate respective program embeddings for further demonstrating the effectiveness of different function embeddings to program embeddings. Specifically, to avoid biases, we employ Mean Pooling (see Sec. \ref{sec:experimentsetup}) as the program embedding generation approaches for conducting the ablation study. \keenhash{}-stru and \keenhash{}-sem-based approaches are excluded since jTrans and CLAP do not support aligning both source and binary functions within the same space, and they model binary functions based on assembly code, which differs from that of \keenhash{}. We perform the program clone search where the test dataset of BinaryCorp is randomly divided into repository and query datasets in an 8:2 ratio. Each query binary has at least one same-class sample in the repository. The metrics include mAP@$100$ and mP@$100$. Table \ref{tab:programclonesearchfunctionembeddingmodel} reveals that the Mean Pooling based on \keenhash{} outperforms those based on jTrans and CLAP. For example, \keenhash{}-Mean Pooling achieves 0.8445 in mAP@100, outperforming jTrans by 21.24\%, and CLAP by 1.05\%. This outcome indicates that a more effective function embedding model can potentially enhance the effectiveness of generated program embeddings. Furthermore, our function embedding model encodes source and pseudo functions into the same space, supporting both \keenhash{}-stru and sem methods where the former is better in code obfuscation scenarios (Sec. \ref{sec:eval:againstobfuscation}), while the latter is more effective in huge code reuse ones (Sec. \ref{sec:eval:programclonesearch}).

\section{\keenhash{} Discussion}

\begin{table}[t]
    \centering
    \caption{\revision{Performance of fine-tuned function embedding models across different base LLMs with different sizes.}}
\scalebox{0.75}{\begin{tabular}{c|c|c|c}
\toprule
\revision{\textbf{Fine-tuned Base LLM}} & \revision{\textbf{MRR}} & \revision{\textbf{Recall@1}} & \revision{\textbf{Recall@5}} \\
\midrule

\revision{\code{StarCoder-1B}~\cite{starcoder1b}} & \revision{0.8588} & \revision{0.7932} & \revision{0.9291} \\
\revision{\code{Pythia-1B}~\cite{pythia1b}}    & \revision{0.8572} & \revision{0.7975} & \revision{0.9286} \\
\revision{\code{Pythia-410M}~\cite{pythia410m}}  & \revision{0.8523} & \revision{0.7915} & \revision{0.9267} \\
\revision{\code{Pythia-160M}~\cite{pythia160m}}  & \revision{0.7958} & \revision{0.7257} & \revision{0.8793} \\
\revision{\code{Jina-137M}~\cite{jina137m}}    & \revision{0.7723} & \revision{0.6947} & \revision{0.8651} \\

\bottomrule
\end{tabular}}
\label{tab:functionembeddingmodelfinetune}
\end{table}

\noindent \textbf{\revision{Comparison with Different Sizes of LLMs.}} \revision{As mentioned in Sec. \ref{sec:functionembeddinggeneration}, we fine-tune the \code{Pythia-410M} to have our function embedding model. However, LLMs with larger model sizes are generally expected to improve their comprehension. Here, we demonstrate that \code{Pythia-410M} achieves the best balance between performance and resource consumption. Therefore, we select it for our paper. The training, validation, and test datasets are split from the one introduced in Sec \ref{sec:dataset}, with a ratio of 8:1:1. The various pre-trained base LLMs for fine-tuning, across different sizes, are selected with their popularity and can be found in Table \ref{tab:functionembeddingmodelfinetune}. We use pseudo functions to retrieve source functions to evaluate the quality of both modalities. The evaluation metrics include MRR, Recall@1, and Recall@5, following the ones used in previous works~\cite{wang2022jtrans, wang2024clap}. As shown in Table \ref{tab:functionembeddingmodelfinetune}, \code{Pythia-410M} has a similar performance compared with \code{Pythia-1B}~\cite{pythia1b} and \code{StarCoder-1B}~\cite{starcoder1b}. The larger-scale LLMs have almost no performance improvement, while they will cost more resources including: computing consumption, time, and video memory. Compared with \code{Pythia-160M}~\cite{pythia160m} and \code{Jina-137M}~\cite{jina137m}, \code{Pythia-410M} significantly outperforms both of them. Therefore, we consider \code{Pythia-410M} to be the optimal among these LLMs, due to its performance and resource consumption.}

\noindent \textbf{\revision{Combination of \keenhash{}-stru and sem.}} \revision{\keenhash{}-stru and sem represent a binary program from two different perspectives (Sec. \ref{sec:programembeddinggeneration}), each with its own advantages (Sec. \ref{sec:eval:programclonesearch} and \ref{sec:eval:againstobfuscation}). Combining their respective advantages may further enhance the performance and robustness of \keenhash{}. A potentially direct strategy is to use hybrid search~\cite{hybridsearch}. It conducts respective \keenhash{}-stru and sem similarity evaluations simultaneously, and merges and reranks/reweights the two sets of (paired) results based on normalized similarity scores. We leverage hybrid search on \keenhash{} for program clone search against the IoT dataset, with the same settings in Sec. \ref{sec:eval:programclonesearch}. The importance (i.e., weight) of \keenhash{}-stru and sem is set to 0.5 each. Our experimental results show that the hybrid search achieves 0.9384 mAP@100 and 0.9300 mP@100. The direct combining strategy does not exhibit significant improvement compared to \keenhash{}-stru and sem (see Table \ref{tab:programclonesearch}). We plan to explore how to effectively combine these two types of program embeddings in the future.}

\noindent \textbf{\revision{Fragility of NoS.}} \revision{\keenhash{}-sem leverages NoS feature (see Sec. \ref{sec:semanticsbasedembeddinggeneration}) as one of the factors for weighting function embeddings and performs well shown in our experiments. However, NoS is expected to be fragile against the string \code{libcall} expansion with known small lengths~\cite{llvm, gcc}. Thus, it may vary with compilation options (e.g., O0 vs. O3), affecting weighting results. In addition, real-world malware also usually obfuscates, hides, and avoids string literals in functions (e.g.,  tshd~\cite{tshd} and Zygug~\cite{zygug} in MLWMC~\cite{dambra2023decoding}), which can render NoS less effective. To mitigate this issue, we also introduce the LoC feature as another factor to enhance the robustness of \keenhash{}-sem (see Table~\ref{tab:programclonesearch}).}